\documentclass[10pt]{article}
\usepackage{graphicx}
\usepackage{amssymb,amsmath}
\usepackage{float}
\usepackage{tcolorbox}
\usepackage{amsthm}
\usepackage{graphicx}
\usepackage{hyperref}
\hypersetup{
	colorlinks=true,
	linkcolor=blue,
	filecolor=magenta,      
	urlcolor=cyan}

\begin{document}

\title{Circle packing in regular polygons}
\author{Paolo Amore \\
\small Facultad de Ciencias, CUICBAS, Universidad de Colima,\\
\small Bernal D\'{i}az del Castillo 340, Colima, Colima, Mexico \\
\small paolo@ucol.mx}

\maketitle

\begin{abstract}
We study the packing of a large number of congruent and non--overlapping circles inside a regular polygon. We have devised  efficient algorithms that allow one to generate configurations of $N$ densely packed circles inside a regular polygon and we have carried out  intensive numerical experiments spanning several polygons (the largest number of sides considered here being $16$) and 
up to $200$ circles ($400$ circles in the special cases of the equilateral triangle and the regular hexagon) . 
Some of the configurations that we have found possibly are not global maxima of the packing fraction, particularly for $N \gg 1$, 
due to the great computational complexity of the problem, but nonetheless they should provide good lower bounds for the packing fraction at a given $N$.
This is the first systematic numerical study of packing in regular polygons, which previously had only been carried out for the equilateral triangle, the square and the circle.
\end{abstract}

\maketitle
\section{Introduction}
\label{sec:intro}

Circle packing is possibly the prototype of a multidisciplinary problem: for physicists working in soft condensed matter circle packing, or more generally sphere packing, is relevant in the study of systems with a large number of particles interacting via contact (short--range) interactions (the case of particles of fixed irregular non--congruent shapes is equally if not more interesting); for mathematicians, circle packing falls in a wider class of problems, exemplified by the renowned Kepler's conjecture, regarding optimal sphere packing in three dimensional space; 
to computer scientists it represents a demanding computational problem, which can be used as natural testing ground for developing efficient algorithms. Packing is also an important problem in everyday life and even in the organization of natural systems: the Tammes problem, which amounts essentially to circle packing on the surface of sphere, was originally introduced to describe the distribution of pores on pollen grains~\cite{Tammes30,Aste08}.

In general, these problems are very intuitive and grasping their essence does not require particular mathematical abilities; unfortunately they are also very  difficult to solve, exactly or even approximately (numerically). This is exemplified by Kepler's conjecture, which regards sphere-packing in the three--dimensional Euclidean space, and has been proved
only recently by Thomas Hales~\cite{Hales05}, almost $400$ years after its initial formulation.

The relevant quantity that one tries to maximize in packing problem is the packing fraction, defined as the ratio between the 
area covered by the disks (or whatever shape one is working with) and the total area of the container. 
For the case of congruent circles on the infinite plane it has been proved that the maximum value that one can obtain is~\cite{Thue,FejesToth42}
\begin{equation}
\rho_{\rm plane} = \pi/\sqrt{12} \ .
\label{eqrhoplane}
\end{equation}

Clearly $\rho_{\rm plane}$ is also an upper bound to the packing fraction of $N$ congruent disks in a finite domain on the plane: in general, the 
optimal hexagonal packing in the infinite plane cannot be achieved in a finite domain, because of the frustration introduced by the 
borders, which may lower considerably the maximal density~\footnote{For the equilateral triangle and the regular hexagon the hexagonal packing can be realized
for specific values of $N$. Even in this case, however, the maximal density is lower than (\ref{eqrhoplane}) for finite $N$ due to the effect of the border.}. The effect of the border becomes negligible in the limit $N \rightarrow \infty$  and the optimal bound (\ref{eqrhoplane}) can be recovered in this limit.

The main domains in the plane where the packing of congruent disks has been studied before are the circle~\cite{Kravitz67,Reis75,Melissen94b,Graham98,Graham97a,Graham97b,Fodor99,Fodor00,Fodor03}, the equilateral triangle~\cite{Oler61,Melissen93,Melissen94,Melissen95,Payan97,Graham04,Joos21}, the square (see refs.~\cite{Schaer65,Schaer65b,Goldberg70,Goldberg71,Nurmela97,Nurmela99,Nurmela99b,Szabo07,Szabo07b,Markot21,Amore21}) and rectangles of different
proportions~\cite{Melissen97,Graham03,Graham09,Specht10}.  Additionally, results for specific shapes of the container and of the ojbects (which can also be non--congruent) 
can be found in the repositories ~\cite{FriedmanRepo} and  \cite{SpechtRepo}. In particular the "packomania" repository~\cite{SpechtRepo}, curated by E. Specht, contains a incredibly rich collection of configurations (and diagrams), which have recently been extended to include more general regular polygons for modest numbers of disks. 
We also mention the repository \cite{crcpam} which contains the configurations found in \cite{Amore21} for circle packing in the square. 
For the reader interested in obtaining a global picture of the problem ref.~\cite{Hifi09} provides a review on the status of circle and sphere packing.

The main purpose of the present article is to discuss the packing of congruent circles inside domains with the shape of a regular polygon. To achieve this goal, the algorithm 
of ref.~\cite{Amore21} (which in turn is an improvement of the algorithm initially proposed by Nurmela and  Östergård ~\cite{Nurmela97}) is extended and modified. 
The algorithms that we introduce in this work are then used to explore dense configurations of $N$ congruent disks in regular polygons with different number of sides, hopefully 
corresponding, in some cases, to global maxima of the packing fraction (due to the rapidly increasing complexity of the problem as more and more circles are considered,  finding a global maximum of the density becomes extremely difficult even for not so large configurations). 

The paper is organized as follows: in section \ref{sec:method} we describe the various algorithms that we have developed; in section \ref{sec:euler} we discuss Euler's theorem of 
topology for the domains under consideration, in terms of topological charges; in section \ref{sec:bounds} we provide upper bounds for the maximal packing density in domains 
with the shape of a regular polygon; in section \ref{sec:results} we report the numerical results 
and discuss the main features of the configurations obtained; finally, in section \ref{sec:conclusions} we draw our conclusions, highlight the main achievements and discuss possible directions for future work.

\section{The method}
\label{sec:method}

In this section we describe the extension of the methods recently introduced in \cite{Amore21} for the case of a square container to containers with the shape of regular polygons. Following \cite{Nurmela97,Amore21} we need to dispose $N$ points inside  a regular polygon with $\sigma$ sides and arrange them in a way such that the minimal  distance between any two points is maximal. If we take this distance to be twice the radius of the $N$ congruent disks with the centers at the $N$ points, the resulting configuration represents an optimal packing of $N$ disks inside the regular polygon.

We parametrize the coordinates of a point inside a regular polygon with $\sigma$ sides  as
\begin{equation} 
\begin{split} 
x(t,u,\sigma) &= \sin^2 t \ X(u,0,\sigma) \\
y(t,u,\sigma) &= \sin^2 t \ Y(u,0,\sigma)   \ , 
\label{eq_poly_par}
\end{split}
\end{equation}
where $P = (X(u,0,\sigma),Y(u,0,\sigma))$ is a point on the border of the polygon  and 
\begin{equation} 
\begin{split} 
X(u,\delta,\sigma) &= \Gamma(u,\delta,\sigma)  \cos (u) \\
Y(u,\delta,\sigma) &= \Gamma(u,\delta,\sigma)  \sin (u) \ ,
\end{split}
\end{equation}
with
\begin{equation}
\Gamma(u,\delta,\sigma) \equiv \left(\delta +\cos \left(\frac{\pi }{\sigma}\right)\right) \sec
\left(\frac{\pi }{\sigma}-(u \bmod \frac{2 \pi }{\sigma})\right) \ .
\end{equation}

\begin{figure}[H]
	\begin{center}
		\bigskip\bigskip\bigskip
		\includegraphics[width=5cm]{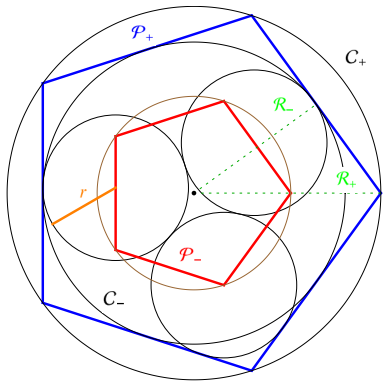}\hspace{1cm}
		\bigskip
		\caption{Optimal configuration for 3 disks inside a pentagon}
		\label{Fig_poly}
	\end{center}
\end{figure}

In this parametrization the coordinates of the centers of the disks are allowed to vary inside a regular polygon of unit circumradius (which corresponds to using $\delta = 0$
in the formulas above), whereas 
\begin{equation}
\begin{split}
R_-  &= r + \cos \frac{\pi}{\sigma} \\
R_+  &= 1 + r \sec \frac{\pi}{\sigma} 
\end{split}
\end{equation}
are the circumradius and the apothem of the regular polygon containing all the disks. This situation is illustrated in Fig.~\ref{Fig_poly} for the case of a regular pentagon ($\sigma=5$): the inner (red) pentagon is the region where the points (centers of the disks) are allowed to access, whereas the outer (blue) pentagon is the region where 
the disks can move.

The perimeter and area of the outer polygon are 
\begin{equation}
\begin{split}
\mathcal{P}(r,\sigma) &= 2 \sigma  \left(r  \tan \left(\frac{\pi }{\sigma }\right)+\sin \left(\frac{\pi }{\sigma }\right)\right) \\
\mathcal{A}(r,\sigma) &= \frac{\sigma}{2} \   \sin \left(\frac{2 \pi }{\sigma }\right) \left(1+r  \sec \left(\frac{\pi }{\sigma }\right)\right)^2 \ .
\end{split}
\end{equation}

Similarly the perimeter of the inner polygon is 
\begin{equation}
\begin{split}
\mathcal{P}_-(r,\sigma) &= 2 \sigma  \sin \left(\frac{\pi }{\sigma }\right)  \ .
\end{split}
\end{equation}

We define the packing fraction as the ratio between the area inside the polygon occupied by the disks and the total area of the polygon
\begin{equation}
\rho \equiv \frac{N \pi r^2}{\mathcal{A}(r,\sigma)} =  \frac{N \pi r^2 \cot \left(\frac{\pi }{\sigma }\right)}{\sigma  \left(r+\cos
	\left(\frac{\pi }{\sigma }\right)\right)^2} \ .
\end{equation}

As the maximum density that can be achieved in the infinite plane is given by (\ref{eqrhoplane}), the condition $\rho \leq \rho_{plane}$ constrains the dimensions of the disks
\begin{equation}
0\leq r \leq \varsigma(N,\sigma) \equiv \frac{\sigma  \cos \left(\frac{\pi }{\sigma }\right)}{ \sqrt{2 \sqrt{3} N \sigma  \cot \left(\frac{\pi }{\sigma }\right)}-\sigma } \ .
\end{equation}
In particular, for $N \gg 1$ 
\begin{equation}
\varsigma(N,\sigma) \approx \sqrt{\frac{\sigma  \sin \left(\frac{2 \pi }{\sigma }\right)}{4 \sqrt{3} }}  \frac{1}{\sqrt{N}} + O\left[\frac{1}{N}\right] \ .
\label{eq_varsigma}
\end{equation}

It is natural to introduce the packing efficiency given by
\begin{equation}
\varepsilon \equiv \frac{r}{\varsigma(N,\sigma)} \leq 1 \ ,
\end{equation}
where $\varepsilon = 1$ corresponds to reaching the highest possible density at the given $N$.

Similarly we define
\begin{equation}
\xi \equiv \frac{N_b r}{\mathcal{P}_-} \leq 1 \ ,
\end{equation}
representing the fraction of the perimeter of the inner polygon covered by the peripheral 
disks ($N_b$ is the number of such disks). In this case $\xi =1$ does not necessarily correspond to reaching the maximal density (for instance, if one arranges $n^2$ disks in a square packing inside a square, then $\xi = 1$, but the density is not optimal).

In addition to these parameters, we also introduce 
\begin{equation}
\nu \equiv \frac{N_v}{\sigma}
\end{equation}
where $N_v$ is the number of vertices of the regular polygon which are occupied by a disk, and
\begin{equation}
\delta \equiv \frac{L}{2 N r}-1  \geq 0 
\end{equation}
where $L$ is the minimal length of a path that goes through all the points crossing them only once (i.e. the minimal path in the associate travelling salesman problem). Configurations of maximal density with $\delta=0$, if found, correspond to the packing of a {\sl necklace}.  A necessary condition such that the a configuratiion of $N$ points is a necklace is that each point  is at a minimal  distance from an even number of two other points. The packing of filaments insides finite regions has interesting 
application in physics and biophyisics (see refs.~\cite{Brito04,Boue07,Gomes08,Stoop08,Gomes10,Bayart11,Oskolkov11,Guven12,Deboeuf13,Gomes15,Grossman21}).

Before describing the algorithms that we have introduced to obtain high--density packing configurations, it is useful to 
specify three possible conventions in the problem setup:
\begin{itemize}
	\item {\rm I}: the regular polygon has an apothem $R_+ = 1+ r \sec \frac{\pi}{\sigma}$, where $r$ is the radius of the disks (this corresponds to the case represented in Fig.~\ref{Fig_poly});
	\item {\rm II}: the apothem of the regular polygon is held fixed at $R_+ = 1$;
	\item {\rm III}: the circle of the disks is held fixed at $r=1/2$;
\end{itemize}

In Table \ref{table_1} we report the main properties of the polygons using the three conventions.

\begin{table}
\caption{Different conventions for the packing of circles inside regular polygons. The results are expressed either using the  radius of the disks in {\rm I} or {\rm II}. }
\begin{center}
{\begin{tabular}{@{}|c|c|c|c|@{}} 
\hline
& ${\rm I}$ & ${\rm II}$ & ${\rm III}$  \\
\hline 
$r$ & $r$ & $\frac{r}{1+r  \sec \left(\frac{\pi }{\sigma }\right)}$ & $\frac{1}{2}$ \\
& $\frac{r}{1-r \sec \left(\frac{\pi }{\sigma }\right)}$ & $r$ & $\frac{1}{2}$\\
\hline
$R_+$ & $1 + r \sec \frac{\pi}{\sigma}$ &  $1$  &  $\frac{1 + r \sec \frac{\pi}{\sigma}}{2r}$ \\
& $\frac{1}{1-r \sec \left(\frac{\pi }{\sigma }\right)}$ &  $1$  &  $\frac{1}{2 r}$ \\
\hline
$\mathcal{A}$ &  $\frac{\sigma}{2} \   \sin \left(\frac{2 \pi }{\sigma }\right) \left(1+r  \sec \left(\frac{\pi }{\sigma }\right)\right)^2$ & $\frac{\sigma}{2} \  \sin \left(\frac{2 \pi }{\sigma }\right)$ &  $\frac{\sigma}{8r^2} \   \sin \left(\frac{2 \pi }{\sigma }\right)$ \\
& $\frac{\sigma  \sin \left(\frac{2 \pi }{\sigma }\right)}{2 \left(r \sec \left(\frac{\pi }{\sigma }\right)-1\right)^2}$ & $\frac{\sigma}{2} \  \sin \left(\frac{2 \pi }{\sigma }\right)$ &   $\frac{\sigma  \tan \left(\frac{\pi }{\sigma }\right) \left(r-\cos \left(\frac{\pi }{\sigma }\right)\right)^2}{4 r^2}$  \\
\hline
$\mathcal{P}$ &  $2 \sigma  \left(r  \tan \left(\frac{\pi }{\sigma }\right)+\sin \left(\frac{\pi }{\sigma }\right)\right)$ & $2 \sigma  \sin \left(\frac{\pi }{\sigma }\right)$ &  $\frac{\sigma}{r}  \sin \left(\frac{\pi }{\sigma }\right)$ \\
& $\frac{2 \sigma  \sin \left(\frac{\pi }{\sigma }\right)}{1-r \sec \left(\frac{\pi }{\sigma }\right)}$ & $2 \sigma  \sin \left(\frac{\pi }{\sigma }\right)$ &
$\frac{\sigma  \tan \left(\frac{\pi }{\sigma }\right) \left(\cos\left(\frac{\pi }{\sigma }\right)-r\right)}{r}$ \\ 
\hline
\end{tabular}	\label{table_1}}
\end{center}
\end{table}

Notice that $\rho$ and $\xi$ defined earlier are {\sl scale invariant} quantities and therefore their expressions do not depend on the convention used.

We can now briefly describe the algorithms used in this paper.

\subsection{Algorithm 1}

This algorithm has been recently introduced in Ref.~\cite{Amore21} as an improvement of an algorithm 
proposed earlier by Nurmela and Östergård in \cite{Nurmela97}. The main difference of the present implementation from the one of  \cite{Amore21} is the different parametrization of the domain. Refs.~\cite{Nurmela97,Amore21} are limited to the square: in that case a point inside the square is parametrized as
\begin{equation}
(x,y) = \frac{\ell}{2} (\sin t,\sin u) \ \ \ , \ \ \  -\frac{\pi}{2} \leq  t,u \leq \frac{\pi}{2}
\label{eq_para_NO}
\end{equation}

Observe that this parametrization is fundamentally different from the one of eq.~(\ref{eq_poly_par}), in
a way that is analogous to the relation between  cartesian and polar coordinates. 
In particular, using eq.~(\ref{eq_poly_par}) one is able to describe the entire border of the domain in terms of a single angle ($0 \leq u \leq 2 \pi$).

Once a parametrization is chosen (in this case eq.~(\ref{eq_poly_par}) ) the original  
constrained optimization problem  is converted to an unconstrained one, which is easier to deal with.

As we have done in \cite{Amore21}, we then consider a energy functional
\begin{equation}
V =  \sum_{i=2}^N \sum_{j=1}^{i-1} \left(\frac{\lambda}{r_{ij}^2}\right)^s \mathcal{F}_{ij}(\epsilon,\alpha) \ ,
\label{eq_energy}
\end{equation}
where
\begin{equation}
\mathcal{F}_{ij}(\epsilon,\alpha) =  \left[\left( \cos^2 t_i + \epsilon \right)
\left( \cos^2 t_j + \epsilon \right) \right]^\alpha 
\label{eq_F}
\end{equation}
and $\epsilon >0$ and $\alpha \leq 0$.

The term $\mathcal{F}_{ij}(\epsilon,\alpha)$ provides a border repulsion, which is essential to avoid 
that excessive number of points reach the border in the earlier stages of the algorithm. As we have remarked in \cite{Amore21}, once a point is deposited on the border, it is practically impossible to remove it and therefore the algorithm will fail to produce a sufficiently dense configuration if the optimal configuration has less border points. On the other hand,  if we set 
$\alpha=0$ we have $\mathcal{F} = 1$ (this limit would be the analogous of \cite{Nurmela97}, with a different parametrization) and the border repulsion disappears.

Although this aspect of the present implementation is similar to that of \cite{Amore21}, there is a fundamental difference in the explicit form of $\mathcal{F}$, which reflects the different 
parametrization used.  Clearly there is considerable freedom in choosing the functional form of $\mathcal{F}$, as long as it avoids the points to fall on the border too early. Eq.~(\ref{eq_F}),
while not unique, it is very simple, thus limiting the amount of numerical load.

The parameter $\lambda$ in eq.~(\ref{eq_energy}) plays an important role: although the potential is always repulsive, the intensity of the repulsion is much stronger for  $r^2 \leq \lambda$, while dying off for $r\rightarrow \infty$. Moreover  the repulsion is essentially confined to the $r^2 \leq \lambda$
for $s \gg 1$. The appropriate choice is to set $\lambda$ to the minimal distance between any two points in the set: in this way, for large $s$, the interaction mimics the contact interaction between rigid disks.

The pseudocode for this algorithm is essentially the pseudocode of \cite{Amore21} (with minor differences) that we report in the appendix for completeness.

As we have noticed above, what makes our algorithm 1 more effective than the original algorithm of ref.~\cite{Nurmela97} in producing dense configurations is the ability to avoid the overpopulation of the border in the early stages of the algorithm, where the interaction is long range.  The border repulsion that we have introduced limits  very much this problem and the efficiency of the algorithm is greatly enhanced.

We have also modified our Algorithm 1 to carry out the minimization of the energy at each value of $s$ using the basin hopping (BH) algorithm. The basin hopping method~\cite{Li87,Wales97,Wales99,Wales03} is considered the standard algorithm for global optimization of large scale numerical problems. When applied to the Thomson problem, for example, it has led to consistent improvement of previous records (see for instance \cite{Wales06}).

In the case of circle packing, however, the search for the global maximum of the density is pursued indirectly by progressively changing the potential from  long range ($s$ small) to short  range ($s \gg 1$) and by  minimizing the energy at each step. Moreover, the energy functional of eq.~(\ref{eq_energy})  is changing  even at a fixed $s$ because of the dependence on $\lambda$.
For this reason one may expect that the effectiveness of the method could be affected.

In our numerical explorations we have found that implementing the use of basin hopping in algorithm 1 does not lead to 
a noticeable improvement of performance. We plan to further explore the use of the BH algorithm in packing problem in future work.

\subsection{Algorithm 2}

As we have discussed in the previous section, the fundamental inspiration for improving the algorithm of \cite{Nurmela97} and thus obtaining algorithm 1, comes from regarding a configuration of disks as a physical systems of repelling "charges" and recognizing that, in a conductor, charges tend to go to the surface (in three dimensions): in two dimensional domains, for long range interactions, the charge density tends to accumulate at the border leaving the central region depleted. The inclusion of border repulsion allows one to limit greatly this behavior and reflects into a larger probability of generating dense configurations.

The configurations obtained with algorithm 1 are typically very dense, but, in many cases, they can be still improved to achieve larger densities. To improve the packing configurations obtained with algorithm 1, we are using algorithm 2, originally proposed in \cite{Amore21}.

As for algorithm 1, this algorithm is inspired by a concrete physical analogy: it is well known from experiments that the dense packing of uniform spheres or solids of different shape can be improved if the container in which the objects are poured is shaken~\cite{Pouliquen97,Kudrolli2010}.

\subsection{Algorithm 3: a variance minimizing algorithm}

Although the algorithms of \cite{Amore21} allow one to obtain dense configurations of congruent circles inside a regular polygon, these configurations can still be improved by solving the set of coupled nonlinear equations corresponding to the contacts between different neighboring circles and between the peripheral circles and the border.  This is the approach followed by Nurmela and Östergård in \cite{Nurmela97}. 

This procedure may appear straightforward, but its implementation is not completely trivial; in some cases, what is expected to be a contact it is in reality a "false contact", meaning that two disks are extremely close without touching each other. 
In such case the corresponding equation should be eliminated by the set. Unfortunately one cannot know a priori whether a contact is "true" or "false" and therefore a considerable amount of experiments might be needed. 

To avoid the complications that we have just described we have introduced a different algorithm that  allows to improve the results obtained with the two main algorithms, without having to solve a system of nonlinear equations. In general we will refer to this process, as well as the procedure described by NÖ,  as a "refinement".

The algorithm is described in the pseudocode in the appendix and it is based on the observation that minimal distances for the circles that are expected to be in contact with other circles in a densely packed configuration obtained with algorithms 1 and 2 (or even different algorithm) will be typically distributed about some certain value with a small variance. The smallness of this variance will be an indication of the "quality" of the configuration. Ideally, if we can slightly move 
the points in a way that the variance vanishes, it means that we have reached a "perfect" packing. Here we are using "perfect" not as a synonym of densest but just to mean that the contacts between the disks are  fully enforced within the numerical precision. While moving the points, however, we have to be careful in holding the border points  on the border, otherwise unwanted solutions could be produced (think for example of the points gradually collapsing into a unique point). 

\begin{figure}[H]
\begin{center}
\bigskip\bigskip\bigskip
\includegraphics[width=6cm]{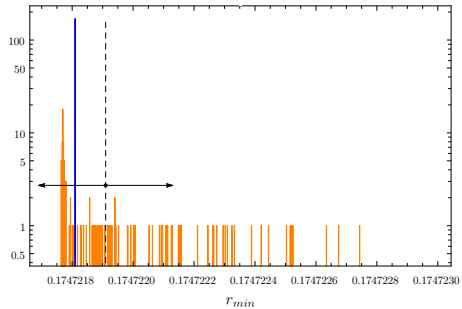} \\		
\caption{Histograms for the distribution of possible contact for $100$ disks inside a pentagon. The orange histogram corresponds to the data obtained using algorithm 1+2, whereas the blue histogram corresponds to applying algorithm 3 to the previous data. The vertical dashed line is the average of the data of the first histogram, whereas the horizontal arrows mark the variance of these data, $\Sigma_1 \approx 2.2 \times 10^{-7}$. The variance of the second set is several orders of magnitude smaller, $\Sigma_2 \approx 5.6 \times 10^{-21}$.}
\label{Fig_2}. 
\end{center}
\end{figure}

Figure \ref{Fig_2} is worth a thousand worth: here we consider the configurations of $100$ disks inside a pentagon, obtained 
using algorithms 1 and 2 (orange plot) and algorithms 1+2+3 (blue plot). The first configuration is very dense, $\rho_1 \approx 0.821430066209$, but the possible contacts are distributed with a rather large variance,  $\Sigma_1 \approx 2.2 \times 10^{-7}$.
After applying algorithm 3, a slightly denser configuration is found, with
$\rho_2 \approx 0.821430442804$, with a variance that is several orders of magnitude smaller, $\Sigma_2 \approx 5.6 \times 10^{-21}$.

Special considerations can be given to packing problems inside the equilateral triangle and the regular hexagons: these are the only two polygons that allow hexagonal packing for specific values of $N$;
based on the experience that we have accumulated in solving packing problems, we have observed  that sometimes the algorithms produce dense configurations of disks with one or more "holes" (i.e. space that could accomodate an extra disk). Sometimes these configurations could correspond to genuine global maxima of the packing fraction~\footnote{For the case of the equilateral triangle with $N=n (n+1)/2$ disks ($n=1,2,\dots$) it is known that  the disks are arranged in a triangular lattice for the optimal configuration;  Oler~\cite{Oler61} has conjectured that the configurations with $N-1$ disks are  obtained from the former by eliminating one of the disks. This conjecture is sometimes also attributed to Erd\"os.}, although more often they could be improved.

In these cases it can be appropriate to follow these simple steps:
\begin{itemize}
\item Identify the "holes" in the configuration at hand (let $N_h$ be the number of holes);
\item Pick $N_h$ disks on the border (either randomly or among the disks with less neighbors) (it may be convenient to select disks that are in contact);
\item Use the $N_h$ disks above to fill the $N_h$ holes previously detected;
\item Apply algorithm 2 to the resulting configuration;
\end{itemize}

\begin{figure}[H]
\begin{center}
\bigskip\bigskip\bigskip
\includegraphics[width=4cm]{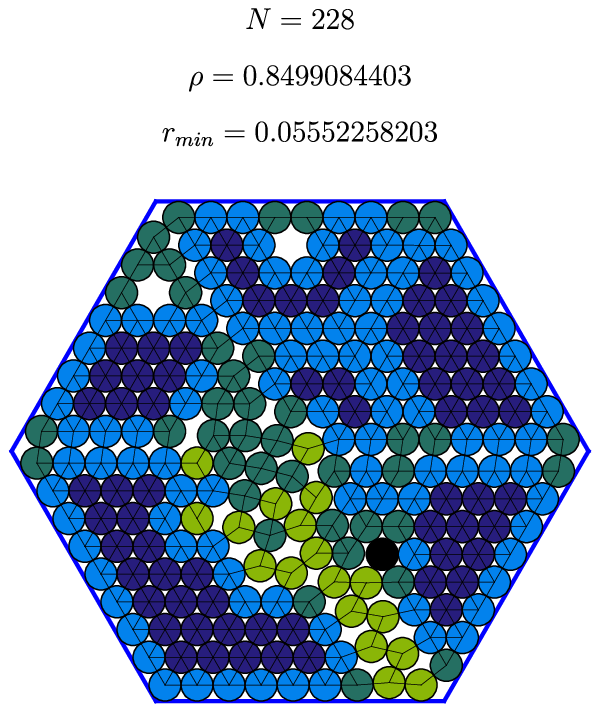} \hspace{1cm}
\includegraphics[width=4cm]{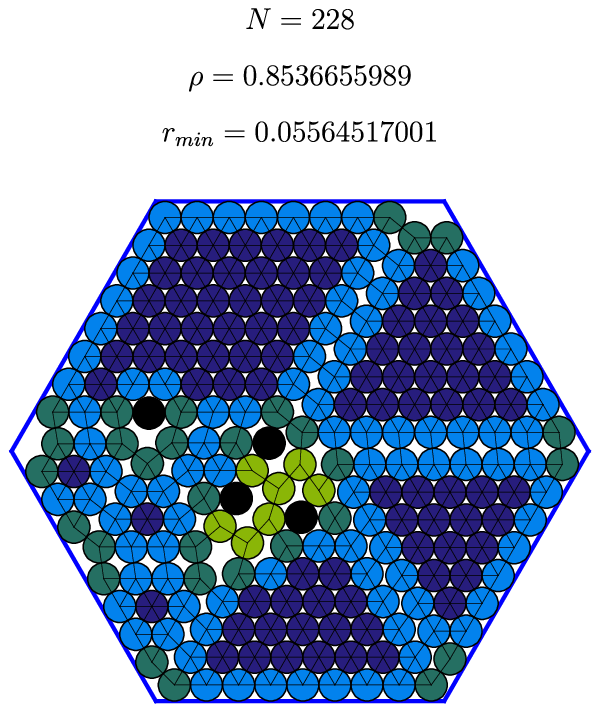} \\		
\caption{Left: Configuration with $228$ disks inside a hexagon, found with Algorithm 1 and 2; Right: Configuration obtained applying Algorithm 4 and 2 to the configuration in the left plot. }
\label{Fig_holes}
\end{center}
\end{figure}

One example of this procedure is given in fig.~\ref{Fig_holes}; the configuration in the right plot is obtained from the configuration in the left plot following the
steps described above.

Similarly, a dense configuration with $N$ disks and $N_h$ holes can be used to produce dense configurations of $N+k$ disks, by filling  $k$ holes and applying algorithm 2.

\begin{figure}[H]
\begin{center}
\includegraphics[width=8cm]{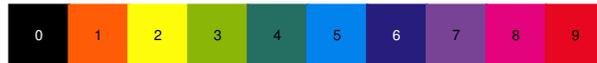}\hspace{1cm}
\caption{Color scheme used to represent the number of contacts (with the border and with other disks) or the number of sides of a Voronoi cell (see the discussion in next section).}
\label{Fig_contacts}
\end{center}
\end{figure}

The color scheme used in this figure is explained in Fig.~\ref{Fig_contacts}: disks that have $n$ contacts (within a given tolerance) with other disks or the walls of the container are represented with a specific color (the same color code will be later used in the Voronoi diagram representations to distinguish cells with different number of sides). 

\section{Euler's theorem and packing}
\label{sec:euler}

In the absence of a border, the densest packing of congruent disks in the plane corresponds to a hexagonal lattice, with a density (packing fraction)
$\rho = \pi/\sqrt{12}$. When borders are added, defining a finite region, this configuration is no longer possible and, as a result, lower densities 
are achieved. The border introduces a {\sl geometrical frustration}, in the sense that the hexagonal structure cannot be attained simultaneously for all the
disks. A similar situation also occurs in the absence of a border, but in presence of curvature, such as  for the 
problem of $N$ charges interacting on  non--planar surfaces (particularly on the sphere)~\cite{Bowick00,Bowick02,Bowick03,Wales06,Wales09,Bowick09,Wales13}. This last problem is usually referred to as "Thomson problem".

To a more fundamental level, Euler's theorem of topology requires  that for any tessellation of the manifold one must obey the equation
\begin{equation}
N_V - N_E + N_F = \chi  \ ,
\label{Euler}
\end{equation}
where $N_V$, $N_E$ and $N_F$ are respectively the number of vertices, edges and faces of the tessellation and $\chi$ is the Euler characteristic ($\chi = 2$ for the sphere). Eq.~(\ref{Euler}) can be cast in an equivalent form in terms of the topological charges, $q_i  = 6 - c_i$, where $c_i$ is the {\sl coordination number } of the $i^{th}$ point, as~\cite{Bowick09}
\begin{equation}
Q = \sum_{i=1}^{N} q_i = 6 \chi \ .
\label{Euler2}
\end{equation}

In particular, since $Q=12$ for a sphere, it is not possible to tile a sphere only with hexagons: this is actually directly observed in the minimal energy 
configurations obtained for the Thomson problem, where the simplest arrangement corresponds to $12$ pentagonal disinclination immersed in a 
"sea" of hexagons (however, as the number of particles grows, new kind of defects, which still comply with Euler's theorem appear).

\begin{figure}[h]
\begin{center}
\bigskip\bigskip\bigskip
\includegraphics[width=3.5cm]{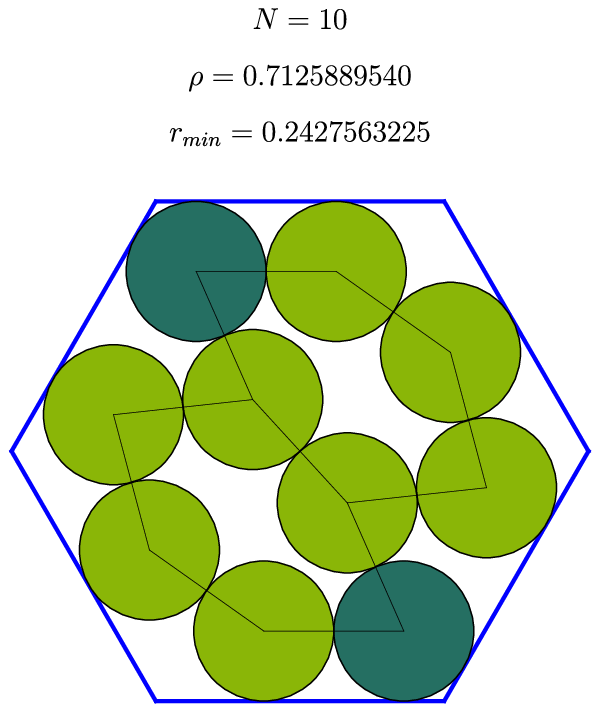}
\hspace{1cm}
\includegraphics[width=3.5cm]{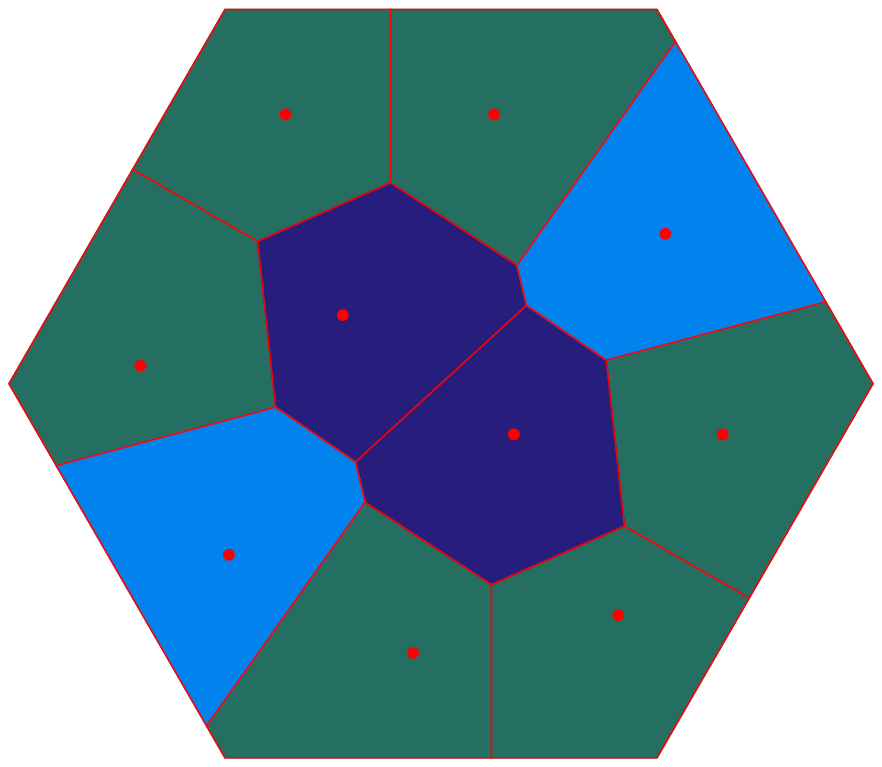}
\bigskip
\caption{Left: Optimal packing of $10$ disks inside an hexagon; Right: Voronoi diagram of the configuration.}
\label{Fig_Euler}
\end{center}
\end{figure}

Consider, for example, the optimal packing of $10$ congruent disks in an hexagon, reported in the left of figs.~\ref{Fig_Euler}; in 
the right figure we find the Voronoi diagram for this configuration, consisting of $6$ quadrilaterals, $2$ pentagons and $2$ hexagons 
(represented in different colors). The reader should not be confused by the fact that some of the cells are classified as quadrilaterals although they appear to be pentagons: when one of the vertices of these cells corresponds to a vertex of the hexagonal domain,  they can be continuously deformed to a straight line, unless the vertex falls on the border between two adjacent cells. We will refer to these vertices as "spurious".

\begin{figure}[H]
\begin{center}
\bigskip\bigskip\bigskip
\includegraphics[width=3.5cm]{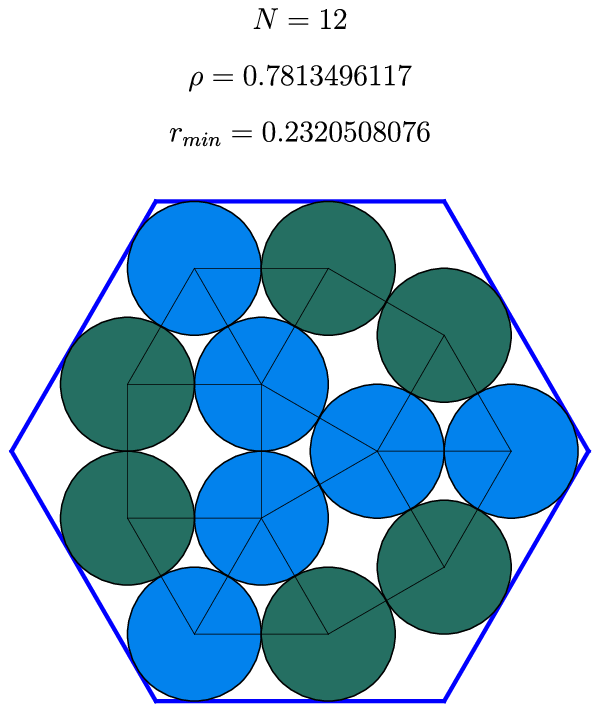}
\hspace{1cm}
\includegraphics[width=3.5cm]{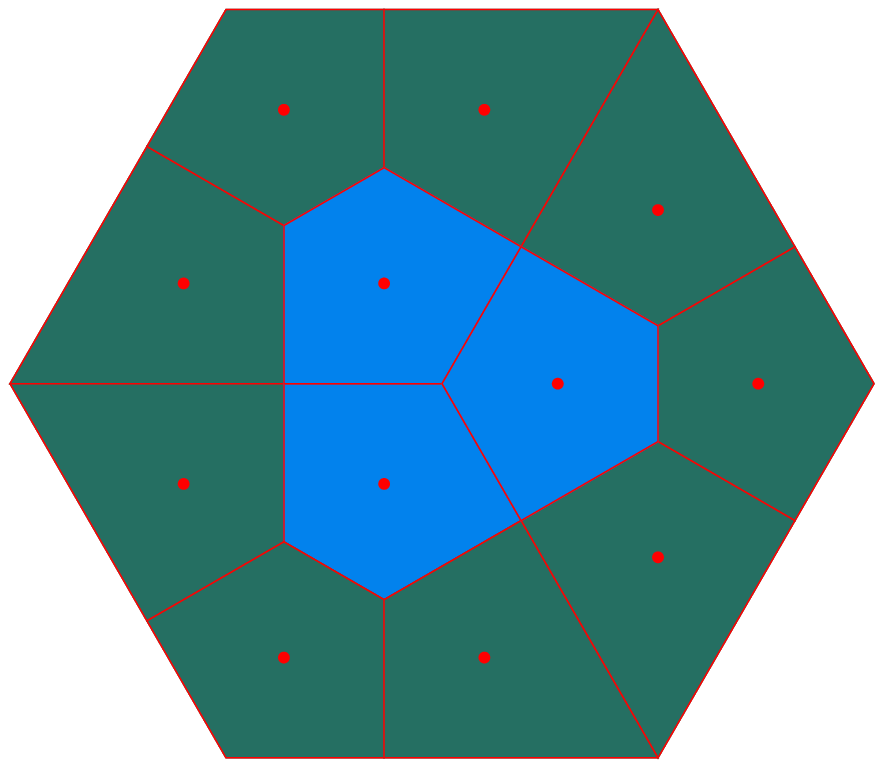}
\bigskip
\caption{Left: Optimal packing of $12$ disks inside an hexagon; Right: Voronoi diagram of the configuration.}
\label{Fig_Euler2}
\end{center}
\end{figure}

\begin{figure}[H]
\begin{center}
\bigskip\bigskip\bigskip
\includegraphics[width=3.5cm]{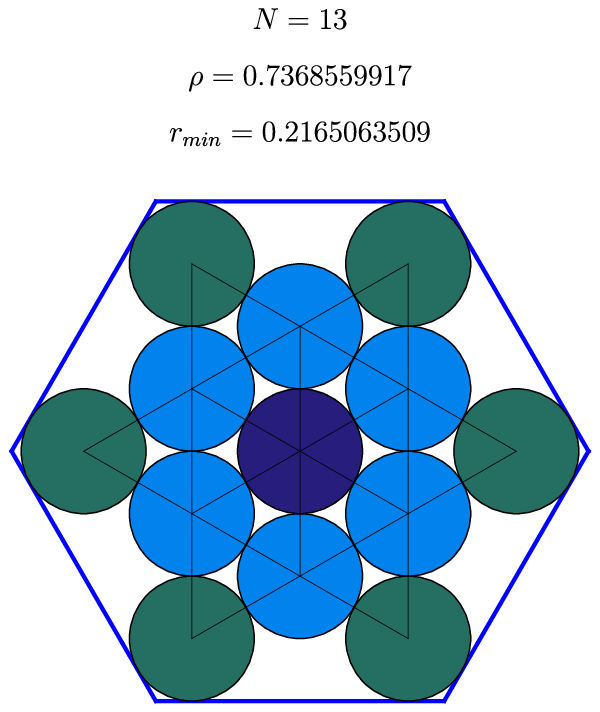}\hspace{2cm}
\includegraphics[width=3.5cm]{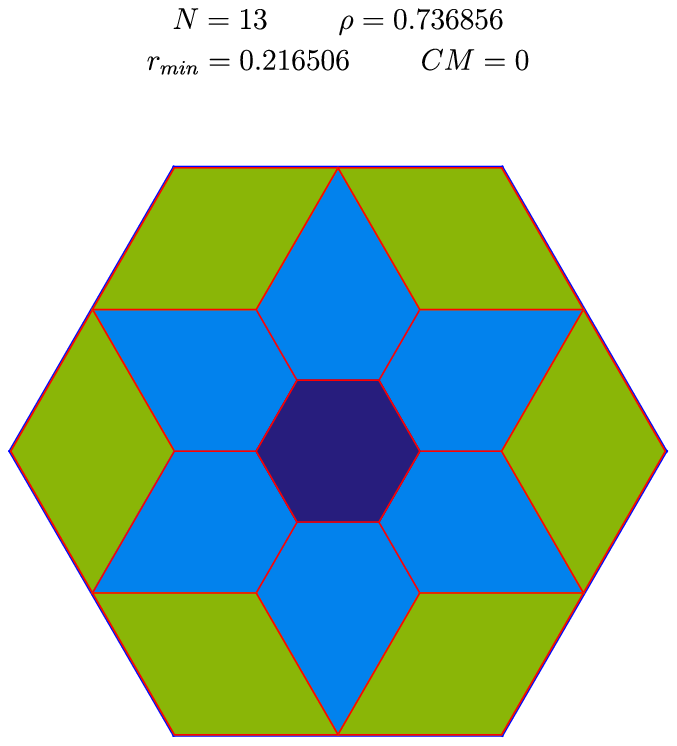}\\
\bigskip
\caption{Best packing for $N=13$ equal circles inside a hexagon.}
\label{Fig_Euler2b}
\end{center}
\end{figure}

\begin{figure}[H]
\begin{center}
\bigskip\bigskip\bigskip
\includegraphics[width=3.5cm]{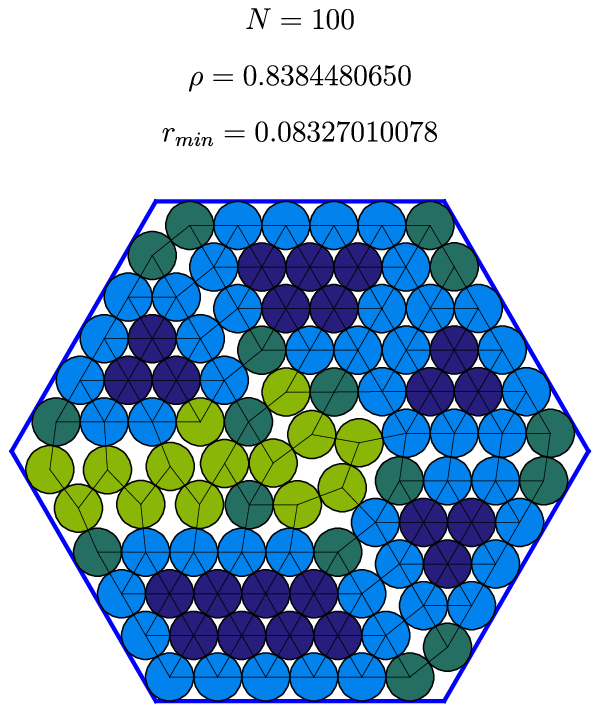}
\hspace{1cm}
\includegraphics[width=3.5cm]{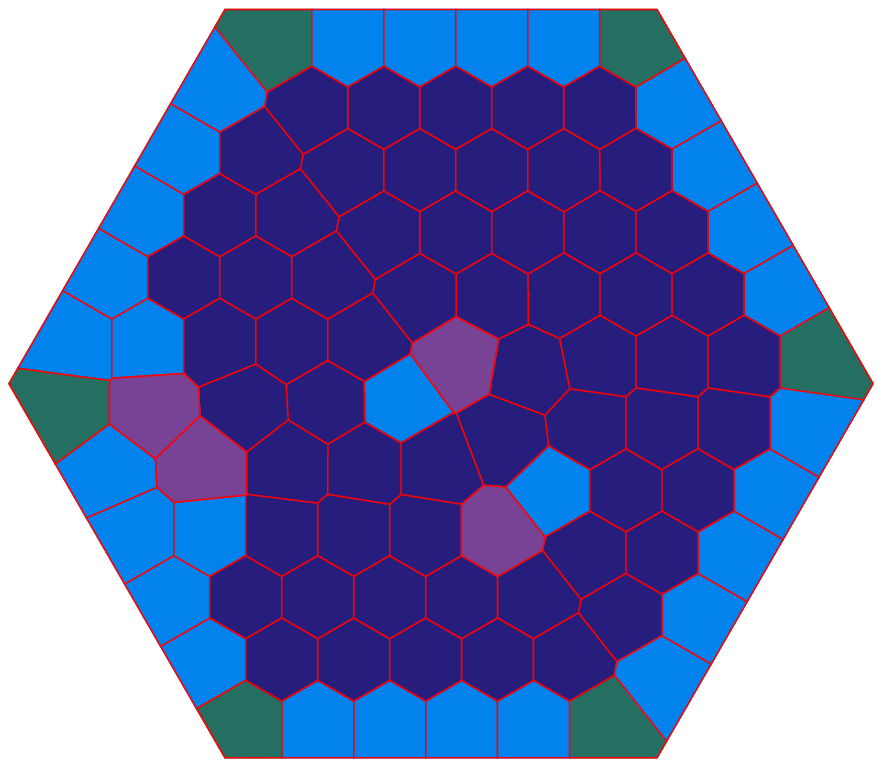}
\bigskip
\caption{Left: Optimal packing of $100$ disks inside an hexagon; Right: Voronoi diagram of the configuration.}
\label{Fig_Euler3}
\end{center}
\end{figure}

Looking at the figures we see that we have $N_b = 8$ cells on the border of the hexagon (corresponding to the number of peripheral disks)
(and $\bar{N}_s = 6$ spurious vertices). With some imagination we can project this diagram onto a sphere (for example, using a stereographic projection from the plane to the sphere) and thus put ourselves in the canonical situation of the Thomson problem. When we perform this operation, the region of the plane external to the hexagon is mapped to a polygonal region on the sphere with exactly $N_b$ vertices (in this case an octagon).

By taking into account the topological charge of the outer face we have
\begin{equation}
Q = \sum_{i=1}^N q_i + (6-N_b) = 12 \ .
\label{Euler3}
\end{equation}

Using eq.(\ref{Euler3}) we have
\begin{equation}
Q = 2 \cdot 1 + 6 \cdot 2 + 2 \cdot 0 + 1 \cdot (-2) = 12 \ .
\end{equation}

Let us now consider the optimal configuration of $12$ disks inside the hexagon for which eq.~(\ref{Euler3})  fails. The straightforward application of eq.~(\ref{Euler2}) in this case  yields
\begin{equation}
Q = 3 \cdot 1 + 9 \cdot 2 + 1 \cdot (-3) = 18 \neq 12
\end{equation}
although one can easily check that Euler's theorem, stated in terms of vertices, edges and faces holds (as it should!). The breakdown of 
eq.~(\ref{Euler3}) has a simple explanation (and equally simple cure): the Voronoi diagram in this case contains three vertices of order four, while 
the discussion in \cite{Bowick09} assumes that only $3$ lines emanate from a vertex. A 4-vertex can however be slightly modified to a pair of 3-vertices, which implies that 2 of the four faces sharing the vertex will gain one edge each. In other words, this is equivalent to 
associate a topological charge of $-2$ to each 4-vertex. A similar argument can be extended to a n-vertex, which acquires a
topological charge $-2 (n-3)$. 

In this way we write
\begin{equation}
Q = \sum_{i=1}^N q_i + \sum_{k} q_k^{(v)}  + (6-N_b) = 12 \ ,
\label{Euler4}
\end{equation}
where $q_k^{(v)}$ is the topological charge on each vertex (for 3-vertices $q^{(v)}=0$).

Applying eq.~(\ref{Euler4}) to the case of Fig.~\ref{Fig_Euler2} we now get the correct answer:
\begin{equation}
Q = 3 \cdot 1 + 9 \cdot 2 + 3 \cdot (-2) + 1 \cdot (-3) =  12  \ . \nonumber 
\end{equation}

It is instructive to look at the configuration of $13$ disks inside the hexagon displayed in Fig.~\ref{Fig_Euler2b}: in this case we have
\begin{equation}
Q = 6 \cdot 3 + 6 \cdot 1 + 6 \cdot (-2) =  12  \ . \nonumber 
\end{equation}

Eq.~(\ref{Euler4}) can be expressed in an equivalent and more practical form
\begin{equation}
\begin{split}
Q &= \sum_{i=1}^{N_{V_{int}}} q_i^{(int)} + \sum_{i=1}^{N_b} (q_i^{(ext)} -1 ) + 6 + \sum_{i=1}^{N_{vert}} q_i^{(v)} \\
&= \sum_{i=1}^{N_{V_{int}}} q_i^{(int)} + \sum_{i=1}^{N_b} \bar{q}_i^{(ext)}  + 6 + \sum_{i=1}^{N_{vert}} q_i^{(v)} = 12 \\
\label{eq_Euler}
\end{split}
\end{equation}
where $\bar{q}_i^{(ext)} \equiv q_i^{(ext)}-1 = 5 - c_i^{(ext)}$ is the topological charge of the $i^{th}$ border cell (in this case the pentagon carries null charge).

A similar formula, but limited to 3-vertices, is reported in refs.~\cite{Bowick07} and \cite{Olvera13}, although in that case the 
topological charge of a border point is reported as $q_i^{(ext)} = 4 - c_i$. This apparent incongruence is just a matter of convention: in our case we are defining
$c_i^{(ext)}$ as the number of sides of the Voronoi cell, which for internal cells is also the number of neighbors; in the case of \cite{Bowick07,Olvera13} $c_i$ is the number of neighbors (observe that for external cells the number of neighbors does not match the number of sides of the cell).

If one eliminates a redundant factor of $6$ from both sides of eq.~(\ref{eq_Euler}) one obtains the simpler form
\begin{equation}
\begin{split}
\sum_{i=1}^{N_{V_{int}}} q_i^{(int)} + \sum_{i=1}^{N_b} \bar{q}_i^{(ext)}   +\sum_{i=1}^{N_{vert}} q_i^{(v)} = 6 \\
\label{eq_Euler_2}
\end{split}
\end{equation}

As we see from eq.~(\ref{eq_Euler_2}) the total topological charge inside the domain must equal $6$; however Euler's theorem does not discriminate among configurations where a number of cells is replaced by the same number of cells (of different shape) carrying the same topological charge.
As the density increases configurations with a complicated defect structure may become  favorable energetically.

\begin{figure}[h]
\begin{center}
\bigskip\bigskip\bigskip
\includegraphics[width=3.5cm]{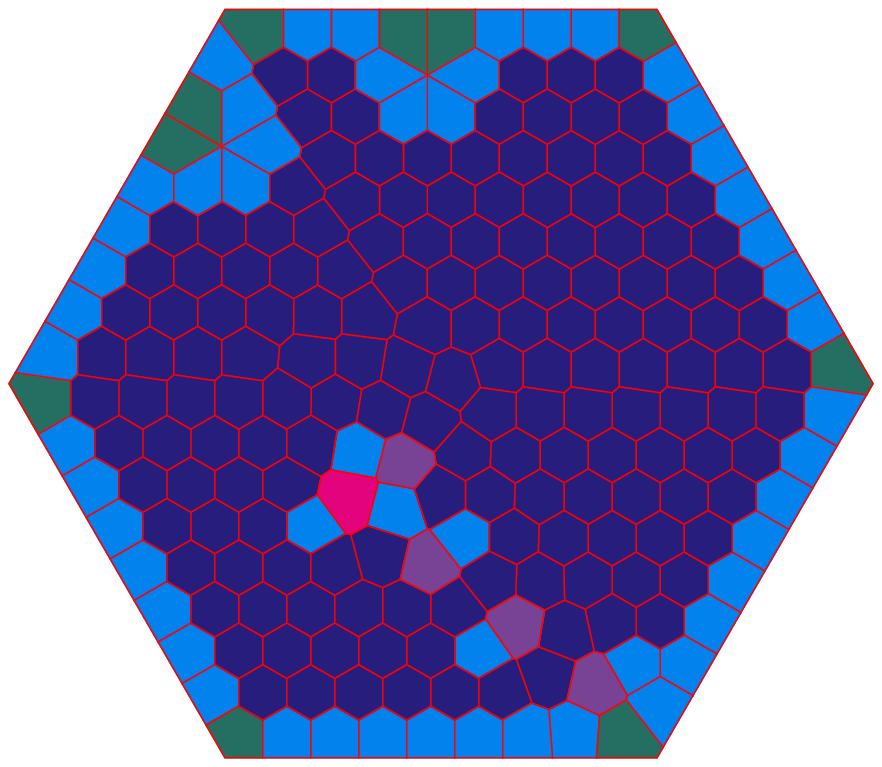} \hspace{1cm}
\includegraphics[width=3.5cm]{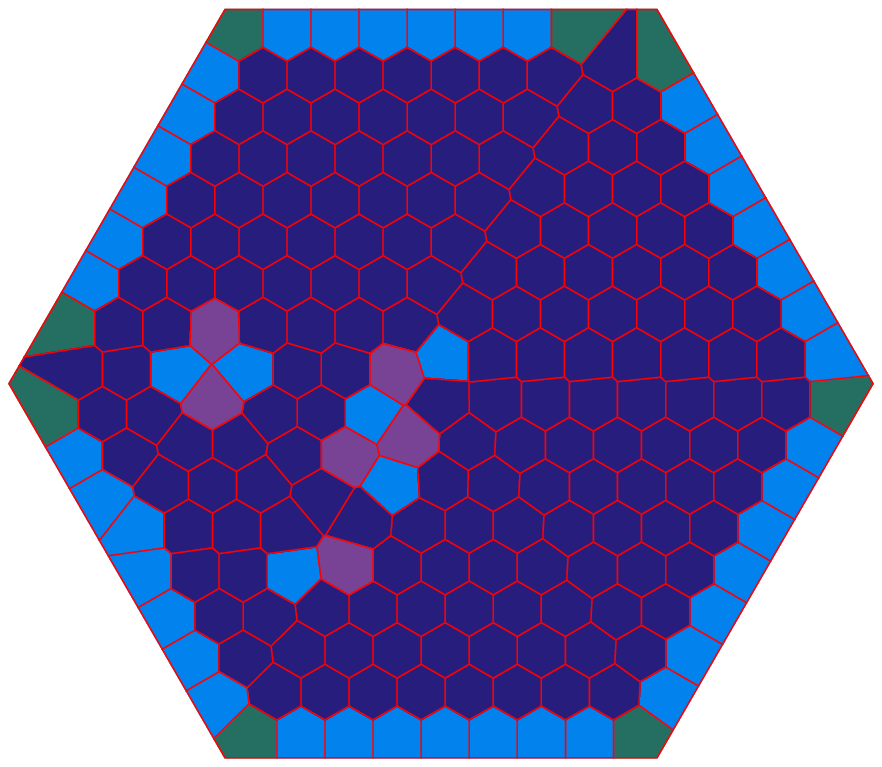} \\		
\caption{Voronoi diagrams for the configurations of Fig.~\ref{Fig_holes}.}
\label{Fig_holes_Voro}. 
\end{center}
\end{figure}

It is worth producing the Voronoi diagrams of the configurations in Fig.~\ref{Fig_holes}, shown in  Fig.~\ref{Fig_holes_Voro}.
The topological charges and vertices for these configurations are reported in  table \ref{table_Voro_228} and one can verify that
Euler's theorem is indeed obeyed ($Q=12$ in both cases). Observe that the configuration on the left contains a larger number of 
defects, as well as two 6-vertices.

\begin{table}[ph]
\caption{Topological charge of the configurations in fig.~\ref{Fig_holes_Voro}. The number of hexagonal cells and 3-vertices is not reported since they don't contribute to the topological charge.}
\begin{center}
{\begin{tabular}{@{}|c|cccc|c|ccc|c|@{}} 
\hline
& \multicolumn{4}{c|}{cells} & $N_b$ & \multicolumn{3}{c|}{vertices} & $Q$ \\
& $4$ & $5$  & $7$ & $8$  &  & $V_4$ & $V_5$ & $V_6$ &  \\
\hline
left of fig.~\ref{Fig_holes_Voro} &  10 &  54 &  4 &  1 & 50 & 0 &0 & 2 & 12 \\
right of fig.~\ref{Fig_holes_Voro} &  8 &  46 &  6 &  0 & 50 & 0 &0 & 0 & 12 \\
\hline
\end{tabular} 	\label{table_Voro_228}}
\end{center}
\end{table}

\section{Upper bounds for the density}
\label{sec:bounds}

Thue's theorem~\cite{Thue} establishes that the densest packing of equal disks in the plane occurs when the disks form a hexagonal lattice. The density that corresponds to this configuration is $\rho=\pi/\sqrt{12}$.  Fejes Toth~\cite{Fejes-Toth50,Fejes-Toth53} has proved that for a given convex domain  $D$ in the plane where $N$ unit disks are packed the following inequality holds
\begin{equation}
N \sqrt{12} < A(D)  \ ,
\label{eq_FT}
\end{equation}
where $A(D)$ is the area of the domain (Segre and Mahler had earlier proved this result for the special case of a convex polygon with at most six sides~\cite{Segre44}).

Groemer~\cite{Groemer60} has sharpened (\ref{eq_FT}) using the results of Segre and Mahler~\cite{Segre44}, obtaining the inequality
\begin{eqnarray}
N \sqrt{12} \leq  A(D) - \varkappa P(D) + \lambda  \ ,
\label{eq_GR}
\end{eqnarray}
where $P(D)$ is the perimeter of the domain and
\begin{equation}
\varkappa \equiv \frac{2-\sqrt{3}}{2} \approx 0.133975 \hspace{.5cm} , \hspace{.5cm} \lambda \equiv \sqrt{12} - \pi (\sqrt{3}-1) \approx 1.1643  \ .
\nonumber
\end{equation}

For a Jordan polygon $\Pi$, Oler~\cite{Oler61} has proved that for $N$ disks of unit diameter
the  tighter inequality inequality holds
\begin{equation}
N \leq \frac{2}{\sqrt{3}} A(\Pi) + \frac{1}{2} P(\Pi) +1   \ .
\end{equation}

We can use the results of Table \ref{table_1} to express Oler's inequality in terms of the disk radius 
for a regular polygon of unit apothem; using this table we have
\begin{equation}
\begin{split}
A(\Pi) &= \frac{\sigma  \tan \left(\frac{\pi }{\sigma }\right) \left(r-\cos \left(\frac{\pi }{\sigma }\right)\right)^2}{4 r^2} \\
P(\Pi) &= \frac{\sigma  \tan \left(\frac{\pi }{\sigma }\right) \left(\cos\left(\frac{\pi }{\sigma }\right)-r\right)}{r}  \ ,
\end{split}
\end{equation}
and
\begin{equation}
N \leq \frac{1}{2\sqrt{3}} \frac{\sigma  \tan \left(\frac{\pi }{\sigma }\right) \left(r-\cos \left(\frac{\pi }{\sigma }\right)\right)^2}{r^2} + \frac{\sigma  \tan \left(\frac{\pi }{\sigma }\right) \left(\cos\left(\frac{\pi }{\sigma }\right)-r\right)}{2r} +1  \ ,
\end{equation}
where $r$ is the disk radius in {\rm II}.

The upper bound on $N$ is reached  for 
\begin{equation}
r = \frac{\sqrt{3} \sqrt{\sigma  \sin \left(\frac{\pi }{\sigma }\right) \left(8 \sqrt{3} (N-1) \cos \left(\frac{\pi }{\sigma }\right)+3 \sigma  \sin \left(\frac{\pi }{\sigma }\right)\right)}+\left(3-2 \sqrt{3}\right) \sigma  \sin \left(\frac{\pi }{\sigma }\right)}{12 (N-1)-2 \left(\sqrt{3}-3\right) \sigma  \tan \left(\frac{\pi}{\sigma }\right)}  \ ,
\label{eq_rmax}
\end{equation}
and  
\begin{equation}
r \approx \frac{\sqrt{\sigma \sin \left(\frac{2 \pi}{\sigma }\right)}}{2 \sqrt[4]{3} \sqrt{n}} - \frac{\left(2 \sqrt{3}-3\right) \sigma  \sin \left(\frac{\pi }{\sigma}\right)}{12 n} + \dots   \ ,
\label{eq_rmax_asym}
\end{equation}
for $N \rightarrow \infty$. Observe that the leading term in eq.~(\ref{eq_rmax_asym}) corresponds to $\varsigma$ of eq.~(\ref{eq_varsigma}).

Using eq.(\ref{eq_rmax}) we obtain an upper bound on the density
\begin{equation}
\begin{split}
\rho &\leq \rho^{(max)}(N,\sigma) = \frac{4 \pi  N}{ \Delta(N,\sigma)  } \\
\end{split}
\label{eq_rhomax}
\end{equation}
where 
\begin{equation}
\begin{split}
\Delta(N,\sigma)  &\equiv -\sqrt{2} \left(\sqrt{3}-2\right) \sec \left(\frac{\pi }{\sigma }\right) \sqrt{\sigma  \sin\left(\frac{2 \pi }{\sigma }\right) \left(8 \sqrt{3} (N-1)+3 \sigma  \tan \left(\frac{\pi }{\sigma}\right)\right)} \nonumber \\
&+ 8 \sqrt{3} (N-1)+2 \left(5-2 \sqrt{3}\right) \sigma  \tan \left(\frac{\pi }{\sigma}\right) \ .
\end{split} 
\end{equation}

For $N \rightarrow \infty$
\begin{equation}
\rho^{(max)}(N,\sigma) \approx \frac{\pi }{2 \sqrt{3}} - 
\frac{\pi}{6} \sqrt{\frac{1}{2} \left(7 \sqrt{3}-12\right)}  
\sqrt{\frac{\sigma  \tan \left(\frac{\pi }{\sigma }\right)}{N}} + \dots  \ .
\label{eq_rhomax_asym}
\end{equation}

G\'asp\'ar and Tarnai~\cite{Gaspar00} have used Groemer's and Oler's inequalities to obtain upper bounds for circle packing in the square, the equilateral triangle and the circle. In particular, the inequalities (6), (14) and (17) of \cite{Gaspar00} for the the equilateral triangle, the square and the circle are 
precisely our inequality (\ref{eq_rhomax}) for $\sigma=3$, $4$ and $\infty$.

Using (\ref{eq_rmax}) we can also obtain an upper bound for  the number of peripheral disks as
\begin{equation}
N_B \leq  \frac{\mathcal{P}(r)}{2r} \ ,
\nonumber
\end{equation}
which reduces to 
\begin{equation}
N_B \lessapprox \sqrt{2 \sqrt{3}\sigma  \tan \left(\frac{\pi}{\sigma}\right) N} 
\label{eq_nb_asym}
\end{equation}
for $N \rightarrow \infty$. Notice that the upper bound is saturated only when the perimeter is completely covered by disks (i.e. $\xi=1$).

\section{Numerical results}
\label{sec:results}

In this section we present the numerical results obtained with the algorithms  described earlier (with the exception of the square, where we use the very precise results of ref.~\cite{SpechtRepo}).  To obtain densely packed configurations of congruent disks inside a domain we have run algorithm 1 a large number of times, followed by algorithm 2 and 3. In general finding the global maximum of the packing fraction of $N$ disks in a given container is quite challenging and rigorous proofs of optimality exist only for special shapes and very limited values of $N$. While we don't expect that all the configurations that we have calculated correspond to global maxima of the packing fraction, we believe that their density should be very close to the maximum. The configurations that we have obtained numerically are available at \cite{crcpam}; plots of these configurations (which include their Voronoi diagrams are available in the supplemental material at \cite{AmoreZenodo})

\begin{figure}[H]
\begin{center}
\bigskip\bigskip\bigskip
\includegraphics[width=8cm]{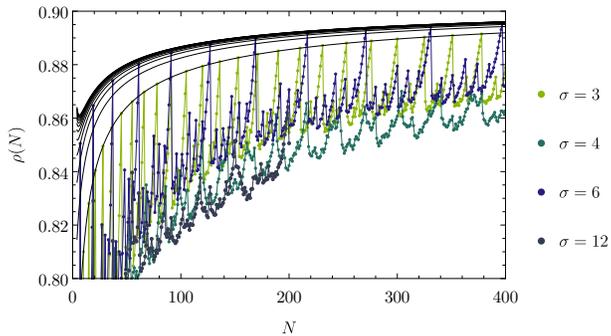}
\caption{Packing fraction for the equilateral triangle, square, hexagon and dodecagon. The solid lines are the upper bound $\rho_{\rm max}(N)$.}
\label{Fig_dens}
\end{center}
\end{figure}

\begin{figure}[H]
\begin{center}
\bigskip\bigskip\bigskip
\includegraphics[width=8cm]{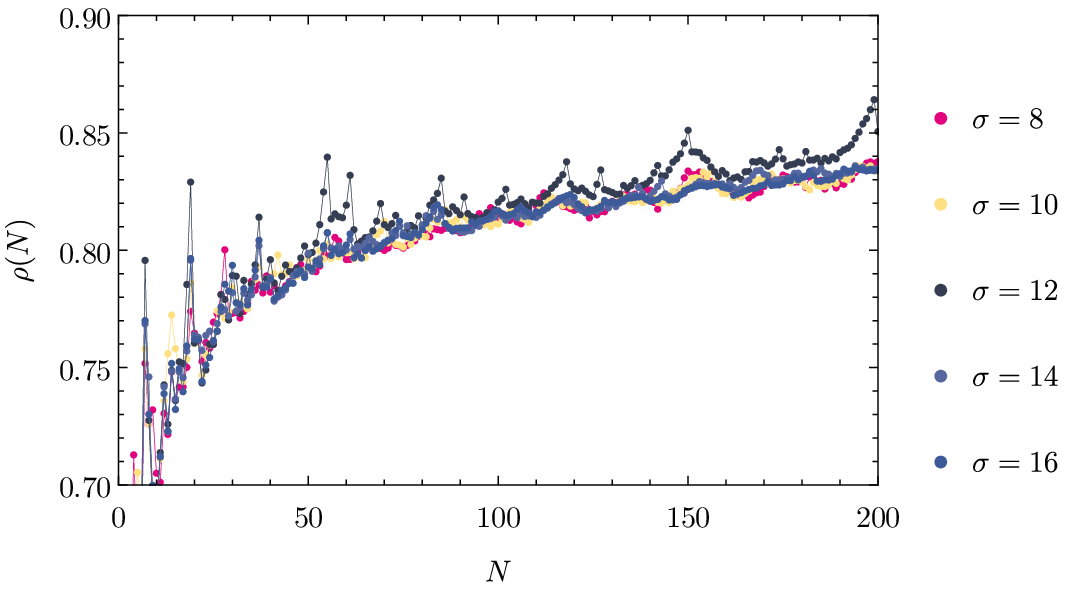}
\caption{Packing fraction for selected regular polygons with an even number of sides.}
\label{Fig_dens2}
\end{center}
\end{figure}

\begin{figure}[H]
\begin{center}
\bigskip\bigskip\bigskip
\includegraphics[width=8cm]{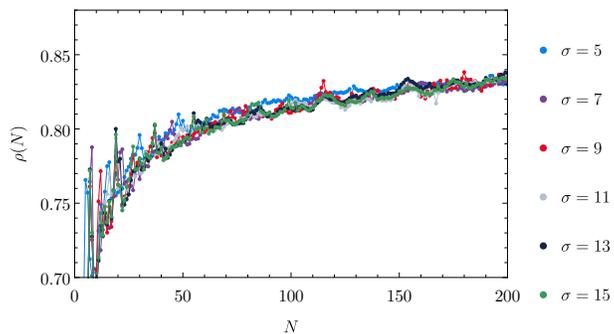}
\caption{Packing fraction for selected regular polygons with an odd number of sides.}
\label{Fig_dens3}
\end{center}
\end{figure}

The most important quantity in our analysis is certainly the packing fraction;  we have plotted it for the several domains considered in this paper
in Figs.~\ref{Fig_dens}, \ref{Fig_dens2} and \ref{Fig_dens3}. Due to the large number of domains that we have considered, we have considered separately 
the  cases of special domains (the equilateral triangle, the square, the hexagon and the dodecagon), of regular polygons with even number of sides and 
regular polygons with odd number of sides, respectively in Figs.~\ref{Fig_dens}, \ref{Fig_dens2} and \ref{Fig_dens3}.

The equilateral triangle and the regular hexagon are special domains, since they admit configurations with triangular packing at specific values of $N$: this is reflected in the large spikes of the density observed in Fig.~\ref{Fig_dens}. At these values, the upper bound $\rho_{\rm max}(N)$ is reached (the continuous curves in the plot represent $\rho_{\rm max})$ for the different domains). The square and the dodecagon also display a pattern of maxima, of more modest value (notice however that the results for the dodecagon cover only $N \leq 200$).

From Figs.\ref{Fig_dens2} and \ref{Fig_dens3}  we see that the behavior of the density does not display in  general large oscillations as found in the previous cases 
(the dodecagon, which is also included in \ref{Fig_dens2}, is seen to outperform the regular polygons with even number of sides included in the plot). Similarly the pentagon, is also performing generally better than the odd regular polygons up to the pentadecagon (with the obvious exclusion of the equilateral triangle), as seen in Fig.~\ref{Fig_dens3}.

For small $N$ the maximum density that can be achieved may be significantly lower than the maximum density in the plane, $\rho_{\rm plane} = \pi/\sqrt{12}$, because
of the frustration introduced by the border. As $N$ gets larger, however, we expect that the packing fraction to increase, as the border effects are increasingly 
less important. In particular,  we must have  $\lim_{N\rightarrow \infty} \rho(N)= \rho_{\rm plane}$. 

We can estimate the size of $N$ congruent disks packed inside a domain of area $\mathcal{A}$, as $r \approx \sqrt{\frac{\mathcal{A}}{N \pi}}$; 
the number of  border disks is then roughly $N_{B} \approx \mathcal{P} \sqrt{\frac{N \pi}{4 \mathcal{A}}}$~\footnote{Strictly speaking $N_B$ is not always well defined, since there may be degenerate configurations with different number of border disks. This happens particularly for the equilateral triangle and the regular hexagon, where the configurations obtained eliminating a disk from a perfectly packed configuration appear to be also optimal. For example in the case of the left plot of Fig.~\ref{Fig_holes}, which represents a non--optimal configuration, the holes could be moved to the border without changing the density. }. On these grounds we expect that the  leading corrections to the packing fraction, at finite $N$, behave as $1/\sqrt{N}$. 

\begin{figure}[H]
\begin{center}
\bigskip\bigskip\bigskip
\includegraphics[width=9cm]{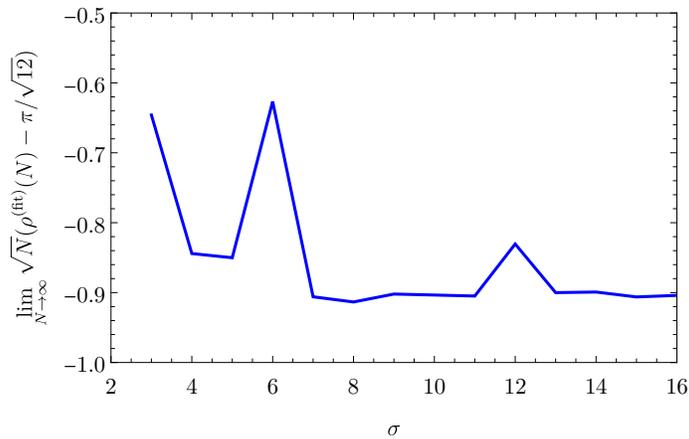}
\caption{$\lim_{N \rightarrow \infty} \sqrt{N} \left( \rho^{(fit)}(N) - \pi/\sqrt{12}\right)$ for different regular polygons as a function of the number of sides.}
\label{Fig_Rho_2}
\end{center}
\end{figure}

Based on this qualitative argument we introduce the fit $\rho^{(fit)}(N) = \frac{\pi}{\sqrt{12}} + \frac{a_1}{\sqrt{N}} + \frac{a_2}{N}$ to describe 
the numerical results of the packing fraction for the regular polygons considered in this paper; 
in Fig.~(\ref{Fig_Rho_2}) we report $a_1$ as a function of $\sigma$ (the number of sides of the polygon). To confirm the special nature of the dodecahedron, 
we see that $a_1$ presents a local maximum at $\sigma=12$, similarly to the cases of the triangle and the hexagon.

\begin{figure}[H]
\begin{center}
\bigskip\bigskip\bigskip
\includegraphics[width=8cm]{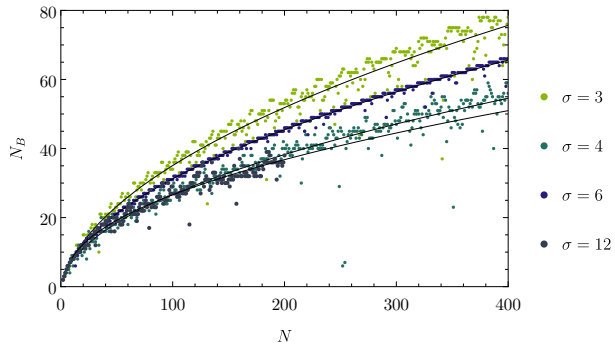}
\caption{Number of peripheral disks, $N_B$, as a function of $N$, for the equilateral triangle, the square, the hexagon and the dodecagon. The solid lines are the fits 
$N_B(N) = \sqrt{N} (a_1 + a_2/\sqrt{N} + a_3/N )$. For the equilateral triangle, square and hexagon the results cover up to $N=400$ (right plot).}
\label{Fig_NB}
\end{center}
\end{figure}

\begin{figure}[H]
\begin{center}
\bigskip\bigskip\bigskip
\includegraphics[width=5.5cm]{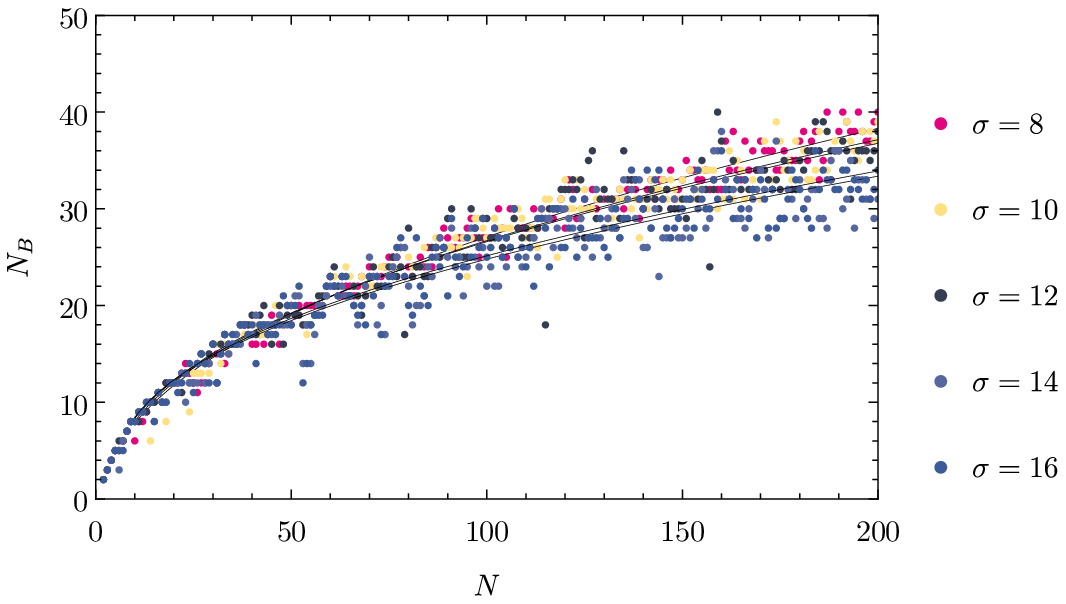} \hspace{0.2cm}
\includegraphics[width=5.5cm]{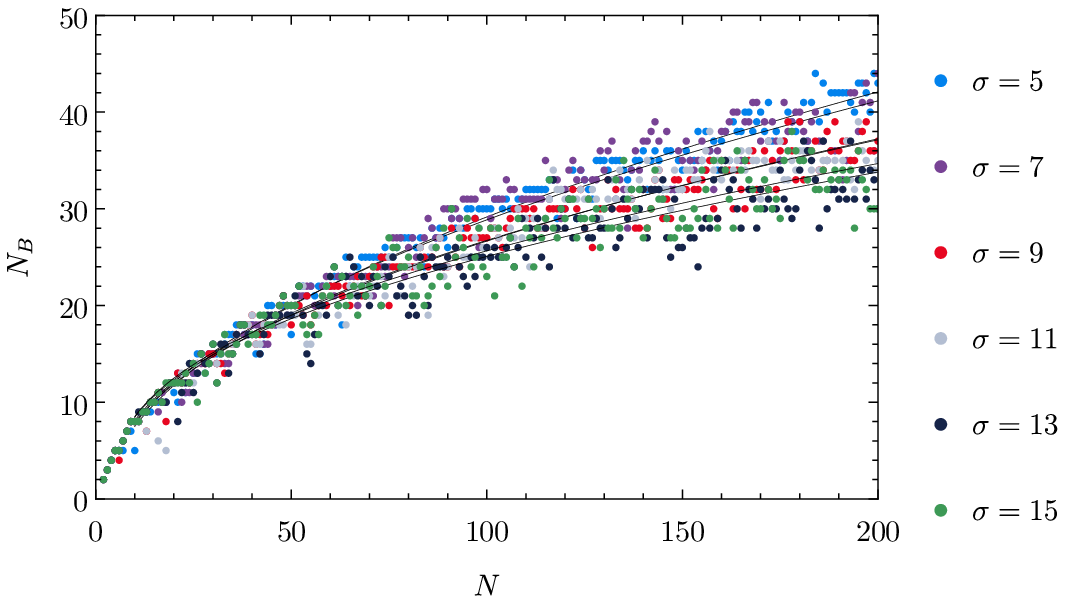}
\caption{Number of peripheral disks, $N_B$, as a function of $N$, for different regular polygons. The solid lines are the fits 
$N_B(N) = \sqrt{N} (a_1 + a_2/\sqrt{N} + a_3/N )$. For the equilateral triangle, square and hexagon the results cover up to $N=400$ (right plot).}
\label{Fig_NB2}
\end{center}
\end{figure}

Our previous qualitative estimate also justifies the fit $N^{\rm (fit)}_B(N) = \sqrt{N} (b_1 + b_2/\sqrt{N} + b_3/N)$ for the number of peripheral disks;
in Fig.~\ref{Fig_NB} and \ref{Fig_NB2} we plot $N_B(N)$, for the different domains that we have studied. It is no surprise that the triangle and the regular 
hexagon allow the largest values of $N_B/N$. The thin solid lines in the plots represent $N^{\rm (fit)}_B(N)$. 
The case of the regular hexagon is particular remarkable since the behavior of $N_B$ vs $N$ is almost monotonically growing.

\begin{figure}[H]
\begin{center}
\bigskip\bigskip\bigskip
\includegraphics[width=7cm]{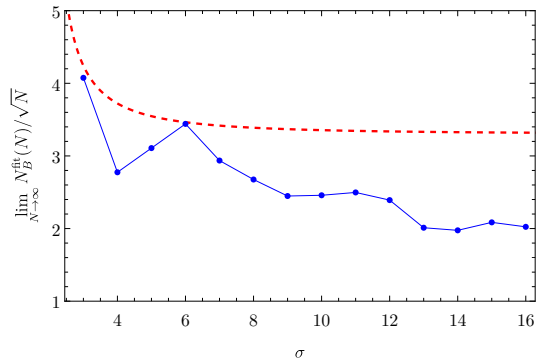}
\caption{$\lim_{N \rightarrow \infty} N_B^{\rm (fit)}(N)/\sqrt{N}$ as a function of the number of sides.  The dashed line is the function $\sqrt{2} \sqrt[4]{3} \sqrt{\sigma  \tan \left(\frac{\pi }{\sigma }\right)}$ from eq.~(\ref{eq_nb_asym}).}
\label{Fig_NB3}
\end{center}
\end{figure}

In Fig.~\ref{Fig_NB3} we plot the coefficient $b_1$ of the fit as a function of $\sigma$ and compare it with the upper bound obtained from  eq.~(\ref{eq_nb_asym}). This bound, as expected, is almost saturated for the hexagon and, to a lesser extent, for the equilateral triangle. 

\begin{figure}[H]
\begin{center}
\bigskip\bigskip\bigskip
\includegraphics[width=9cm]{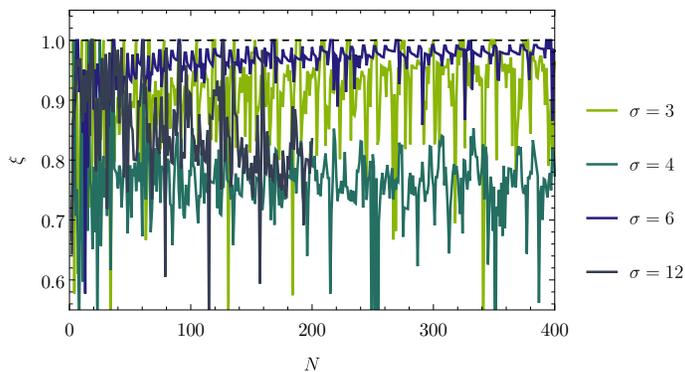}
\caption{Border fraction as a function of $N$ for the equilateral triangle, square, hexagon and dodecagon.}
\label{Fig_Xi}
\end{center}
\end{figure}

\begin{figure}[H]
\begin{center}
\bigskip\bigskip\bigskip
\includegraphics[width=5.5cm]{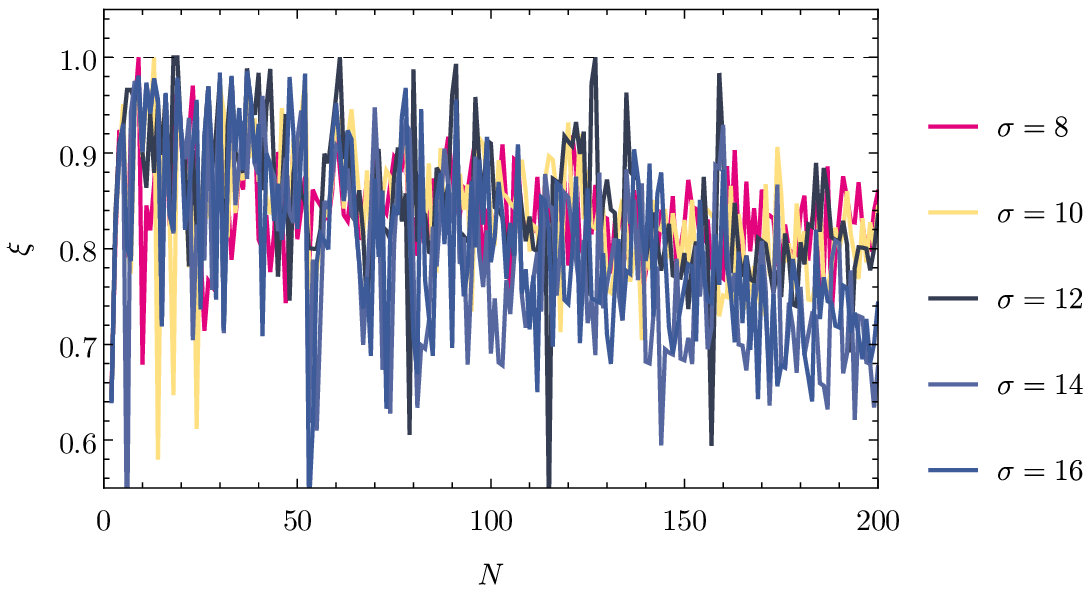} \hspace{0.2cm}
\includegraphics[width=5.5cm]{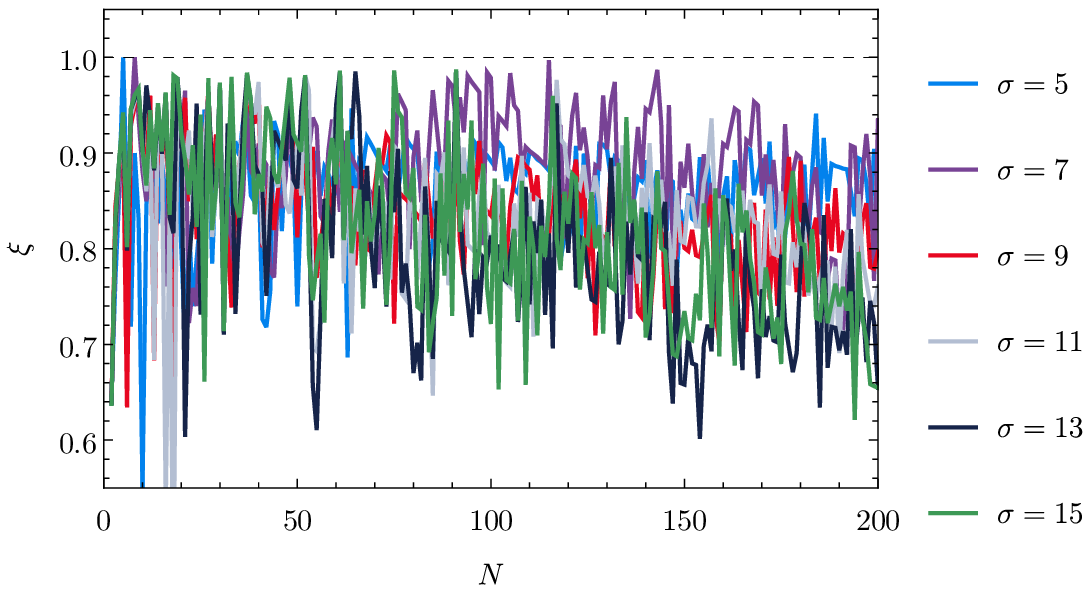}
\caption{Border fraction as a function of $N$ for the equilateral triangle, square, hexagon and dodecagon.}
\label{Fig_Xi2}
\end{center}
\end{figure}

\begin{figure}[H]
\begin{center}
\bigskip\bigskip\bigskip
\includegraphics[width=9cm]{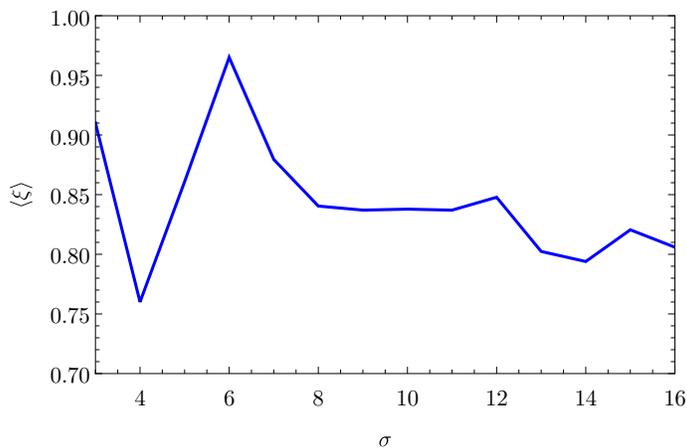}
\caption{$\langle \xi \rangle$ for different regular polygons as a function of the number of sides.}
\label{Fig_Xi3}
\end{center}
\end{figure}

While $b_1$ provides the leading asymptotic behavior of $N_B(N)$ for $N \gg 1$, the border fraction $\xi$ provides information over the border crowding
at finite $N$, as displayed in Figs.~\ref{Fig_Xi} and \ref{Fig_Xi2}.  The largest values of $\xi$ are found for the regular hexagon while 
we find rather small values for the square. The average border fraction, calculated over the set of configurations with 
$2 \leq N \leq 200$ ($400$ for triangle, square and hexagon) is displayed in Fig.~\ref{Fig_Xi3}.

In Table \ref{table_3} we report the configurations of $N$ disks inside a regular polygon ($3 \leq \sigma  \leq 16$) that correspond to
necklaces (in general we have restricted our search to $N \leq 200$, with the exception of  $\sigma=3,4,6$ where $N \leq 400$~\footnote{For the square we have just used the numerical results of \cite{SpechtRepo}}). In our analysis we consider that a configuration corresponds to a necklace if $\delta \leq 10^{-10}$. The regular hexagon and the equilateral triangle possess the largest number of necklaces among the regular polygons studied.

\begin{figure}[H]
\begin{center}
\bigskip\bigskip\bigskip
\includegraphics[width=8cm]{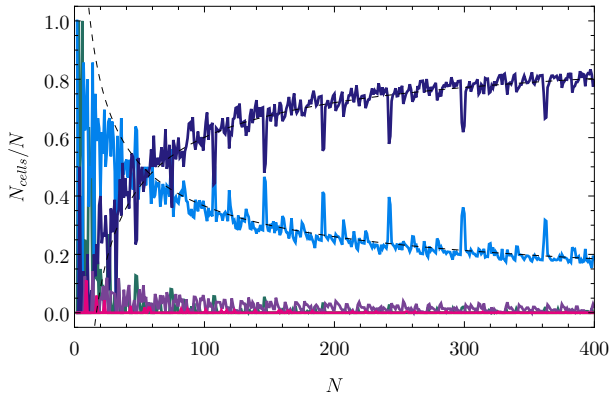}
\caption{Number of Voronoi cells with different number of sides for the regular hexagon. The color scheme corresponds to the one found in fig.~\ref{Fig_contacts}.}
\label{Fig_voro_6}
\end{center}
\end{figure}

The fraction of Voronoi cells of given shape for the packing configurations inside the hexagon is displayed in Fig.~\ref{Fig_voro_6}. The thin dashed lines for the pentagonal and hexagonal cells correspond to the fits $c^{\rm (pentagon)}_1 N^{-1/2} + c_2^{\rm (pentagon)} N^{-1}$ and $1-c_1^{\rm (hexagon)} N^{-1/2} - c^{\rm (hexagon)}_2 N^{-1}$. The qualitative justification for these fits is given by noticing that hexagonal and pentagonal cells dominate at large density and that pentagonal cells are mostly located at the border: since  $N_b$ scales as $\sqrt{N}$ for $N \gg 1$, we have that $N^{\rm (pentagon)}/N  \approx 1/\sqrt{N}$. 
The spikes in the fraction of pentagonal cells (and the corresponding dips in the fraction of hexagonal cells) are due to configurations where the hexagon is "split" into subdomains by straight cuts.

\begin{table}
\caption{Possible necklaces configurations for regular polygons with $3 \leq \sigma \leq 16$ for $N\leq 200$ (with the  exception of $\sigma=3,4,6$ where $N\leq 400$).  }
{\begin{tabular}{@{}|c|c|@{}} 
\hline
$\sigma$			& $N_{\rm necklaces}$  \\
\hline
3  & 2, 3, 6, 9, 10, 12, 14, 15, 20, 21, 24, 25, 26, 27, 28, 31, 35, 36, 39, 40, 42, \\
   & 43, 44, 45, 51, 52, 54, 55, 60, 65, 66, 71, 77, 78, 84, 86, 88, 90, 91, 95, 104, \\
   & 105, 111, 112, 118, 119, 120, 130, 133, 134, 135, 136, 143, 150, 151, 152, 153,  \\
   & 169, 170, 171, 182, 187, 188, 189, 190, 199, 208, 209, 210, 228, 231, 273, 274,  \\
   &275, 276, 300, 310, 321, 324, 349, 350, 351, 374, 377, 378, 389 \\
\hline
4 &  2, 3, 4, 8, 15, 16, 20, 23, 24, 30, 34, 35, 36, 42, 56, 80, 128, 180, 208, 247, 340 \\
\hline
5  & 2, 3, 4, 5, 15, 21, 23 \\
\hline
6  & 2, 3, 4, 5, 6, 7, 10, 11, 12, 14, 16, 17, 18, 19, 26, 27, 29, 30, 33, 34, 35, 36, 37 \\
   & 46, 47, 48, 51, 52, 56, 58, 60, 61, 68, 69, 74, 75, 79, 80, 84, 85, 90, 91, 106, 107, \\ 
   & 108, 114, 118, 119, 120, 123, 124, 126, 127, 146, 147, 153, 154, 160, 161, 167, 168, \\ 
   & 169, 191, 192, 199, 200, 207, 208, 212, 215, 216, 217, 233, 242, 243, 251, 252, 261, \\
   & 269, 270, 271, 289, 298, 299, 300, 309, 310, 319, 320, 324, 325, 329, 330, 331, 350, \\ 
   & 351, 362, 363, 372, 373, 374, 384, 385, 393, 395, 396, 397 \\
\hline
7  & 2, 3, 4, 5, 10, 11, 12 \\ 
\hline
8  & 2, 3, 4, 5, 22, 27, 28, 54, 64 \\
\hline
9  & 2, 3, 4, 5, 7, 10, 11, 37 \\ 
\hline
10 & 2, 3, 4, 5, 10, 12 \\
\hline
11 & 2, 3, 4, 5, 10, 12 \\     
\hline
12 & 2, 3, 4, 5, 6, 7, 10, 11,  12, 18, 19, 31, 37, 47, 61, 73, 91, 127 \\
\hline
13 & 2, 3, 4, 5, 10, 12 \\
\hline
14 & 2, 3, 4, 5, 10 \\ 
\hline
15 & 2, 3, 4, 5, 7, 10, 11, 12, 15, 18, 37, 45, 48 \\
\hline
16 & 2, 3, 4, 5, 10, 12 \\
\hline
\end{tabular} \label{table_3}}
\end{table}

\section{Conclusions}
\label{sec:conclusions}

We have introduced efficient algorithms for calculating configurations of densely packed congruent disks inside regular polygons with an arbitrary number of sides $\sigma$. We have obtained numerical results for $3 \leq \sigma \leq 16$ (with the exception of the square, $\sigma=4$, where we have used the precise configurations reported in \cite{SpechtRepo})  and $N \leq 200$ (for $\sigma=3$ and $\sigma =6$ we have reached $N=400$). Given the computational complexity of the problem we cannot expect that all the configurations that we have obtained are global maxima of the density, but we think that their density should be nearly maximal. The large amount of numerical data has allowed us to perform an analysis of the properties of these systems, which had not been carried out before (as we mention in the introduction, only few regular polygons, such as the square and the equilateral triangle, had been studied in depth in the literature).  Our analysis has focused not only on the packing fraction (density), but also on additional properties of the system that we have identified here: the border fraction, 
for instance, is seen to be quite sensitive to $\sigma$, with the square having the lowest average border fraction of all the polygons that we have studied. Similarly, we have found that, apart from the expected cases $\sigma=3,6$, the dodecagon and the pentadecagon ($\sigma=12,15$) have a larger number of configurations that are necklaces, with respect to the remaining polygons.
Finally, the packing configurations can also be visualized in terms of Voronoi cells, each carrying a topological charge which depends both on the number of sides of the cell and whether the cell is located at the border of the domain or not. The Euler theorem of topology requires that the total topological charge inside the domain add to $Q_{total} = 6$. It is observed that the internal topological charge in these systems tends always to be very small, even at large density, a striking difference
with Coulomb systems where it is observed  to become increasingly negative~\cite{Olvera13}. 
Another important difference with these systems is the presence of higher order vertices, which themselves carry 
a negative topological charge (these vertices are not usually observed in Coulomb systems).

\section*{Acknowledgements}
The research of P.A. was supported by Sistema Nacional de Investigadores (M\'exico). 
The plots in this paper have been plotted using Mathematica~\cite{wolfram} and {\rm MaTeX} \cite{szhorvat}. 
Numerical calculations have been carried out using python ~\cite{python} and numba ~\cite{numba}.

\newpage

\appendix

\section{Pseudocodes of the algorithms}

\begin{tcolorbox}{H}
\begin{center}
\begin{minipage}[c]{0.95\textwidth}			
\begin{center}
{\sl \bf Pseudocode of  algorithm 1 (adapted from \cite{Amore21})}
\end{center}
\begin{itemize}
\item[step 1:]  Generate a random set of N points inside a regular polygon with $\sigma$ sides;
\item[step 2:]  Define the energy functional (6), with $s = s_{in}$ and with $\alpha$ such that $\lim_{s \rightarrow \infty} \alpha = 0$; the initial value of s, $s_{in}$ , may be chosen arbitrarily, but it is convenient that it is not too large, to allow better results;				
\item[step 3:] Minimize the functional in step 2) and use the configuration obtained as new initial configuration, now with $s'= \kappa s$, with $\kappa > 1$;
\item[step 4:] Repeat step 3) until reaching a large value of s (in most of our calculations we have used $s_{fin} = 10^6$ , but larger values might be used if convergence was not reached);
\item[step 5:] Compare the density of the configuration obtained at step 4) with the maximal density for N circles and, if the current density improves the previous record, store the new configuration as the best so far;
\item[step 6:]  Repeat steps 1-5) a large number of times
\end{itemize}			
\end{minipage} 
\end{center}\end{tcolorbox}

\newpage

\begin{tcolorbox}
	\begin{center}
		\begin{minipage}[c]{0.95\textwidth}
			
\begin{center}
{\sl \bf Pseudocode of  algorithm 2 (adapted from \cite{Amore21})}
\end{center}
\begin{itemize}
\item[step 1:] Take a configuration of densely packed disks (for example obtained with Algorithm 1) and perturb randomly the position of the circles;
\item[step 2:] Treat the centers of the circles as point--like particles repelling with an interaction $ \left(\frac{\lambda}{r^2} \right)^{s_{\rm in}}$, with $\lambda = r_{min}^2$, $r_{min}$ being the closest distance between any two points in this set;
\item[step 3:] Minimize the total energy of the system, eq.~(\ref{eq_energy}), for $s=s_{\rm in}$ (typically, at the beginning we choose $s_{\rm in} \approx 10^2-10^3$);
\item[step 4:] Let $s' = \kappa s$, with $\kappa > 1$, and repeat step 3) using the configuration obtained there as the initial configuration;  iterate these steps up to a sufficiently large value of $s$, until convergence has been reached;
\item[step 5:] If the final configuration of step 4) has a higher density of the configuration of 1), use it as new initial configuration and repeat the steps 1--4) as many times as needed, each time updating the initial configuration to be the densest configuration; 
\item[step 6:] If after some iterations the process cannot easily improve the density, modify step 1), by making the amplitude of the random perturbation smaller and repeat the steps 2--5) (in general the initial value of $s_{\rm in}$ will then be taken to be larger);
\item[step 7:]  Stop the algorithm when convergence has been reached (or when the assigned number of iterations has been completed).				
\end{itemize}
\end{minipage} 
\end{center}\end{tcolorbox}

\pagebreak

\begin{tcolorbox}
\begin{center}
\begin{minipage}[c]{0.95\textwidth}

\begin{center}
{\sl \bf Pseudocode of the variance minimizing algorithm}
\end{center}
\begin{itemize}
\item[step 1:]  as initial step in the algorithm, one needs a densely packed configuration of disks inside the polygon, which in our case is obtained using the algorithms 1 and 2;

\item[step 2:] for such a configuration calculate the minimal distance of each point (corresponding to the center of a disk) from any other point in the set and let $d_{min}$ be the smallest of such distances. Circles that are possibly in contact with at least one other circle will have very similar distances. Ideally, if the exact configuration had been reached, these distances would be all the same (assuming that no false contacts are present);
	
\item[step 3:] fix a certain threshold $\eta$ and assume that circles at a distance $d$ such that $0 <d-d_{min}< \eta$ are in contact; identify all pairs of circles that are in contact. We call $p$ the number of such pairs and $\bar{d}_{i}$ (with $i=1,\dots, p$) the distance between the circles forming the $i^{th}$  pair. 
	
\item[step 4:] identify circles that have no contact with other circles;
	
\item[step 5:] identify circles that are in contact with the border; these points are then allowed to move only along the border, but not inward;
	
\item[step 6:] consider the square variance (we work with square distances to avoid dealing with square roots)
\begin{equation}
\Sigma = \frac{1}{p} \sum_{i=1}^p d_i^4 - \left(\frac{1}{p} \sum_{i=1}^p d_i^2 \right)^2 \nonumber
\end{equation}

\item[step 7:] minimize $\Sigma$ allowing the points with at least one contact 	to move in the plane (for circles on the border, only allow movements on the border), while keeping fixed points corresponding to circles with no contacts with other circles; introduce a scale factor $\sigma$ for the 
the amplitude of the movement of the charges (initially one can start with $\sigma  \approx 10^{-4}$);

\item[step 8:] if the density of the new configuration is larger than the initial density then accept  the new configuration and repeat the whole process a
number of times (we normally perform $50$ runs); 

\item[step 9:] if at the end of the whole cycle, the algorithm was not able to find any improvement, then reduce $\sigma$ and $\eta$ 
(typically $\sigma  \rightarrow \sigma/10,\eta \rightarrow \eta/10$) . Repeat until needed.
\end{itemize}
\end{minipage} 
\end{center}
\end{tcolorbox}
\newpage

\section{Generating uniformly distributed random points inside a regular polygon}

The goal is to generate $N$ points uniformly distributed inside a regular polygon. We start by considering the unit
circle ($\sigma = \infty$) and then generalize the discussion to finite values of $\sigma$. Since the probability density for a uniform distribution inside a circle grows linearly as 
\begin{equation}
\rho(r) = 2 r \hspace{1cm} , \hspace{1cm} 0 \leq r \leq 1 ,
\end{equation} 
we see that the fraction of points at a distance $r<1$ goes like $r^2$. However, if we generate the points by using random values of $r$ uniformly distributed on $(0,1)$,  the fraction of points at a distance $r<1$ scales as $r$.
This happens because the points which are close to the center of the circle are generated with a much larger probability, without taking into account the proper scaling of the areas.

However, if the random numbers $\left\{ q_1, q_2, \dots \right\}$ are uniformly distributed between $0$ and $1$, 
the numbers $\left\{ \sqrt{q_1}, \sqrt{q_2}, \dots \right\}$ are distributed linearly, as required by $\rho(r)$. This gives us a simple recipe to generate $N$ points inside a unit circle:
\begin{itemize}
\item[1)] consider $i=1$;
\item[2)] generate a uniform random number $\theta_i$, with $0 \leq \theta_i \leq \pi$;
\item[3)] generate a uniform random number $q_i$, with $0 \leq q_i \leq 1$;
\item[4)] generate the point $P_i \equiv \sqrt{q_i} (\cos\theta_i, \sin\theta_i)$;
\item[5)] $i \rightarrow i+1$ and go back to step 2 until $i>N$.
\end{itemize}

We want now to generalize this result to the case of regular polygons with a unit circumradius: a point inside the polygon is obtained using  eq.~(\ref{eq_poly_par}), which requires specifying the two polar variables $(t, u)$.
Clearly $u$ is the analogous of $\theta$ for the circle and it can be generated using random numbers uniformly distributed on $(0,\pi)$, exactly  as before. It is easy to see that a point with a given angle $u$ must be 
at a distance 
\begin{equation}
r \leq d_{\rm max}(u) = \cos \left(\frac{\pi }{\sigma }\right) \sec \left(\frac{\pi }{\sigma}-(u \bmod \frac{2 \pi }{\sigma })\right)
\end{equation}
to be contained in the regular polygon. If the distance $r$, generated using $\sqrt{q}$ exceeds this bound, the point must be discarded and the operation needs to be repeated until an acceptable value of $r$ is obtained.

\end{document}